\documentclass[12pt]{iopart}
\usepackage{ulem}
\usepackage{graphicx}
\usepackage{color}
\usepackage{dcolumn}% Align table columns on decimal point
\usepackage{enumitem}
\usepackage{bm}% bold math
\usepackage{iopams,graphicx}
\eqnobysec

\newcommand{\cmag}{\color{magenta}}

\def\urt{u({\bf r},t)}

\def\rnqo{\rho({\bf q}=nq_0\hat{z},t)}
\def\pno{\psi_n({\bf q}={\bf 0},t)}

\def\pnrt{\psi_n({\bf r},t)}

\def\urt{u({\bf r},t)}

\def\dut{\langle|\Delta u({\bf r})|^2\rangle}
\def\cqf{C_{uu}^{ET(1)}(\bf{q})}
\def\cqs{C_{uu}^{ET(2)}(\bf{q})}
\def\du{\langle|\Delta u({\bf r})|^2\rangle_}

\newcommand{\cO}{{\cal O}}
\newcommand{\bq}{{{\bf q}}}

\begin{document}

	%%%%%%%%%%%%%%%%%%%%%%%%%%%%%%%%%%%%%%%%%%%%%%%%%%%%%%%%%%%%%%%%%%%%%%%%%%%%%%
	%%								TITLEPAGE

	%title, auther, adress, mail
	\title{Emergent smectic order in simple active particle models}
	\author{Pawel Romanczuk$^{1,2}$
	%\footnote{Current address: Faculty of Life Sciences, Humboldt Universit\"at zu Berlin, 10099 Berlin, Germany}
	, Hugues Chat\'{e}$^{1,3,4}$, Leiming Chen$^{1,5}$, Sandrine Ngo$^{1,3}$ and John Toner$^{1,6}$}

	\address{$^1$ Max Planck Institute for the Physics of Complex Systems, N\"othnitzerstr. 38, 01187 Dresden, Germany}
	\address{$^2$ Faculty of Life Sciences, Humboldt Universit\"at zu Berlin, 10099 Berlin, Germany}
	\address{$^3$ Service de Physique de l'Etat Condens\'e, CEA-Saclay, CNRS UMR 3680, 91191 Gif-sur-Yvette, France}
	\address{$^4$ Beijing Computational Science Research Center, Beijing 100094, China}
	\address{$^5$ College of Science, China University of Mining and Technology, Xuzhou, Jiangsu, 221116, P. R. China}
	\address{$^6$ Department of Physics and Institute for Theoretical  Science, University of Oregon, Eugene, OR 97403, USA}

	\ead{romanczuk@physik.hu-berlin.de}

	%abstract
	\begin{abstract}
		Novel ``smectic-P"  behavior, in which self-propelled particles form rows and move on average along them,  occurs generically 
within the orientationally-ordered phase of simple  models that we simulate. Both apolar (head-tail symmetric) and polar (head-tail asymmetric) models  
with  aligning and  repulsive interactions exhibit slow algebraic decay of smectic order with system size up to some finite length scale, after which faster decay occurs.
In the apolar case, this scale is that of an undulation instability of the rows. In the polar case, this instability is absent, but traveling fluctuations
disrupt the rows in large systems and motion and smectic order may spontaneously globally rotate.
These observations agree with a new hydrodynamic theory which we present here.
Variants of our models also exhibit active smectic ``A" and ``C" order, with motion orthogonal and oblique  to the layers respectively.
		\end{abstract}	
		
	
	\submitto{\NJP}

	\maketitle
	%%%%%%%%%%%%%%%%%%%%%%%%%%%%%%%%%%%%%%%%%%%%%%%%%%%%%%%%%%%%%%%%%%%%%%%%%%%%%%

	%%%%%%%%%%%%%%%%%%%%%%%%%%%%%%%%%%%%%%%%%%%%%%%%%%%%%%%%%%%%%%%%%%%%%%%%%%%%%%
	%%		
	%									MAIN

	\section{Introduction}
%%%%%%%%%%%%%%%%%%%%%% BEGIN MAIN TEXT %%%%%%%%%%%%%%%%%%%%%%%%%%%%%
Active matter, by which we mean  out of equilibrium systems
that locally convert energy into directed motion, 
can potentially exhibit new  phases not present in equilibrium systems.  Even active phases that {\it do} have equilibrium counterparts often behave quite differently from them, in both their fluctuations
\cite{tonertuPRL1995,tonertuPRL1998,tonertuPRE1998,sradititonerEPL2003,vjSCIENCE2007,chatenematicPRL2006,GNF}
and flow properties
\cite{aditisrPRL2002,VoituriezEPL2005,srmadanNJP2007,hatwalnePRL2004,tanniecristinaPRL2003,fieldingPRE2011}.

%{\cred 
All of the work just described and cited dealt with phases
%\sout{Until recently, attention has mainly focused on phases} 
of active particles  
with {\it  orientational} order,   but no {\it translational} order \cite{tonertusrAnnPhy2005,srannurev2010}.
 More recently, attention has turned to translationally ordered phases \cite{bialke_prl_2012,menzel_epl_2012,palacci_science_2013,menzel_pre_2013,redner_prl_2013,menzel_pre_2014,weber_prl_2014}.
 %\sout{This changed with the first theoretical analyses of active matter phases with  translational order}.  
One class of such phases is  ``active smectics" \cite{live-soap, Chen-Toner-2013}, 
%\sout{investigated in those studies} 
which are out of equilibrium analogs of smectic liquid crystal phases.  In these phases, in addition to orientationally  ordering, the particles spontaneously form regularly spaced, liquid-like layers. 
That is, translational symmetry is broken in only {\it one } direction, while the systems themselves are two or three dimensional. 

This theoretical work focused on active analogs of the smectic-A phase, for which particle alignment is perpendicular to the layers. 
These theories \cite{live-soap, Chen-Toner-2013} predicted in particular that smectic order is long-ranged in $d=3$ dimensions and quasi-long ranged  (i.e., power-law correlated) in $d=2$, in strong contrast to the equilibrium case, in which  smectic order is only quasi-long ranged \cite{Caille} in $d=3$ and completely destroyed by thermal fluctuations (i.e., short-range correlated) in $d=2$ \cite{1st}. This phenomenon - that is, activity stabilizing order in low spatial dimensions in which it is absent in equilibrium- is similar to the ``violation" of the Mermin-Wagner theorem that occurs in polar active fluids in two dimensions \cite{tonertuPRL1995,tonertuPRL1998,tonertuPRE1998}.

Prior to the work we report here, none of the predictions made in \cite{live-soap,Chen-Toner-2013} had been confirmed either in experiments or simulations. 
For active particle systems, few striped patterns have been reported so far \cite{wensink_pnas_2012,wensink_jpcm_2012,menzel_epl_2012,menzel_jpcm_2012}, and none have been studied quantitatively as smectic phases.  (See, however, the shaken rice grain experiments of Narayanan et al. \cite{Narayanan-jstat-2006}.)

In this paper we report the generic emergence of active smectic order in simulations of very simple self-propelled particle models, in which the alignment interactions of the familiar Vicsek \cite{vicsek} model are supplemented by short-ranged repulsive interactions between 
particles. Such models have been studied \cite{GNF,gregoire_prl_2004,weber_prl_2014}, 
but not in the high density regimes we consider here.
Simulating both this modified Vicsek model and an apolar (i.e., head-tail symmetric) 
version of it  in $d=2$ space dimensions \cite{chatenematicPRL2006}, 
we observe the development of quasi-long-ranged smectic order over a range of 
length scales in part of the orientationally-ordered phases exhibited by these 
models (cf. Fig. \ref{Fig1-IsoRep}). 
When the repulsion is anisotropic (i.e., dependent on the angle between 
the direction of particle motion and the inter particle relative position vector), we find also
smectic-A (in which the layers are orthogonal to the particle alignment) 
and smectic-C order, in which the layers make an acute angle with the particle alignment (Fig. \ref{Fig2-AniRep} a,b,c). 
Both of these configurations also occur in equilibrium.
But the active smectic configuration we most frequently observe, particularly with isotropic repulsion, is a new state of active matter with no equilibrium counterpart that we call``smectic P", in which the particle alignment is primarily {\it parallel} to the layers (Fig.~\ref{Fig1-IsoRep}b,c). 
A pictorial summary of the polar and apolar smectic A and P phases is given in Fig.~\ref{Fig3-SmPict}.

We deliberately refrain from calling the regions of smectic order we have found {\sl phases}, since in all cases studied, 
quasi-long-ranged smectic order disappears beyond some model-dependent length scale.
This disappearance of order is explained by  comprehensive phenomenological theories of polar and apolar active smectics that we present here. 
These theories predict that symmetry-allowed nonlinearities in the polar case destroy smectic order at the longest length scales, and that 
a long but finite wavelength ``undulation'' instability likewise disorders the apolar case. Both
predictions agree with our simulations.

Finally, we also report a number of spectacular collective dynamical phenomena,  such as spontaneous global rotation of both orientational and smectic order for some polar smectic P, and large-scale swirling structures in polar smectic A, which await theoretical elucidation.

%%%%%%%%%%%%%%%%%%%%%%%%%%%%%%%%%%%%%%%%%%%%%%%%%%%%%%%%%%%%%%%%%
\section{Simulations:  Models and Results}
%%%%%%%%%%%%%%%%%%%%%%%%%%%%%%%%%%%%%%%%%%%%%%%%%%%%%%%%%%%%%%%%%
%%%%%%%%%%%%%%%%%%%%%%%%%%%%%%%%%%%%%%%%%%%%%%%%%%%%%%%%%%%%%%%%%
\subsection{Simulation models}
%%%%%%%%%%%%%%%%%%%%%%%%%%%%%%%%%%%%%%%%%%%%%%%%%%%%%%%%%%%%%%%%%
We consider two classes of models: In  the {\it polar model},  particles 
move along an intrinsic heading {\it vector}, which they align ferromagnetically with that of their 
neighbors. 
In the {\it apolar} model,  particles align their {\it axes} with those of their neighbors, and move 
with equal probability in either direction parallel to this axis.
In both classes, 
we add to this alignment interaction (which is in competition with noise) pairwise repulsion between neighbors. We consider both isotropic repulsion, in  which the repulsion does not depend on the angular position of neighbors with respect to the particle's intrinsic axis, and  anisotropic repulsion, in which it does.

More specifically, in the spirit of the Vicsek model \cite{vicsek}, we consider point particles moving at constant speed $v_0$. 
At discrete timesteps, the position ${\bf r}_i$ of particle $i$ is moved along the unit vector
${\bf u}_i(t+1)=(\cos\theta_i(t+1),\sin\theta_i(t+1))^{\rm T}$:
\begin{eqnarray}
{\bf r}_i(t+1)&=&{\bf r}_i(t)+ v_0 {\bf u}_i(t+1) \;\; {\rm with} \\
\theta_i(t+1) &=& \textnormal{arg}\left[\epsilon(t){\bf A}_i(t) + \beta \, {\bf R}_i(t)\right]+ \sigma \, \chi_i(t)
\label{eq:angle_update}
\end{eqnarray}
where $\chi_i(t)\in[-\frac{\pi}{2}, \frac{\pi}{2}]$ is an angular white noise drawn from a uniform distribution, $\sigma$  is a parameter setting the strength of the angular noise, 
and ${\bf A}_i$ and ${\bf R}_i$ are respectively  ``alignment" and ``repulsion" vectors 
obtained from averages over neighbors defined as those particles located within a distance $r_{int}$., with the default interaction range set to unity (i.e. $r_{int}=1$, see \ref{App:params_repL}). The same unit range was chosen for both interactions for simplicity. 
We have checked that taking different ranges does not qualitatively change the model's behavior.

The two models are distinguished by their alignment interaction:
\begin{equation}
	{\bf A}_i  =  \cases{ 
		{\frac{1}{N_i}} \sum_{j\sim i}{\bf u_j} & \textnormal{polar} \\  
	        {\frac{1}{N_i}} \sum_{j\sim i}\textnormal{sgn}(\cos(\theta_i-\theta_j)){\bf u_j} & \textnormal{nematic}
	   }
\end{equation}
where the $N_i$ neighbors $j$ include particle $i$,
$\epsilon(t)=1$ for the polar model, while for the nematic model $\epsilon(t)$ is randomly chosen to be $\pm 1$ at each step with equal $50\%$ probability. 
The repulsion vector ${\bf R}_i$ is identical for both cases: it is the average over all neighbors of a pairwise force along 
${\bf \hat r}_{ji}$, the unit vector pointing from particle $j$ to $i$. 
The magnitude of this repulsive force is the same for all neighbors in the isotropic case, while, in the anisotropic case, it depends on 
$\phi_{ji}=\textnormal{acos}({\bf u}_i\cdot{\bf \hat r}_{ji})$, the angular position of particle $j$ 
relative to the axis of particle $i$
\begin{equation}
	{\bf R}_i=\cases{ \frac{1}{N_i-1}\sum_{j\sim i,\, i\neq j} {\bf \hat r}_{ji} & \textnormal{isotropic} \\
	\frac{1}{N_i-1}\sum_{j\sim i,\, i\neq j} \cos^2(\phi_{ji}-\gamma) \,{\bf \hat r}_{ji} & \textnormal{anisotropic} %\label{anisorep}
	}
	 \label{isorep}
\end{equation}
Here we only consider $\gamma=0$ and $\gamma=\frac{\pi}{2}$. 
For $\gamma=0$, repulsion is stronger ahead of and behind the particles, 
whereas for $\gamma=\frac{\pi}{2}$ it is stronger towards the left and right.

%%%%%%%%%%%%%%%%%%%%%% FIGURE 1 %%%%%%%%%%%%%%%%%%%%%%%%%%%%%%%%%%%%%
\begin{figure}[tbp!]
	\begin{center}
 	\includegraphics[width=0.8\textwidth]{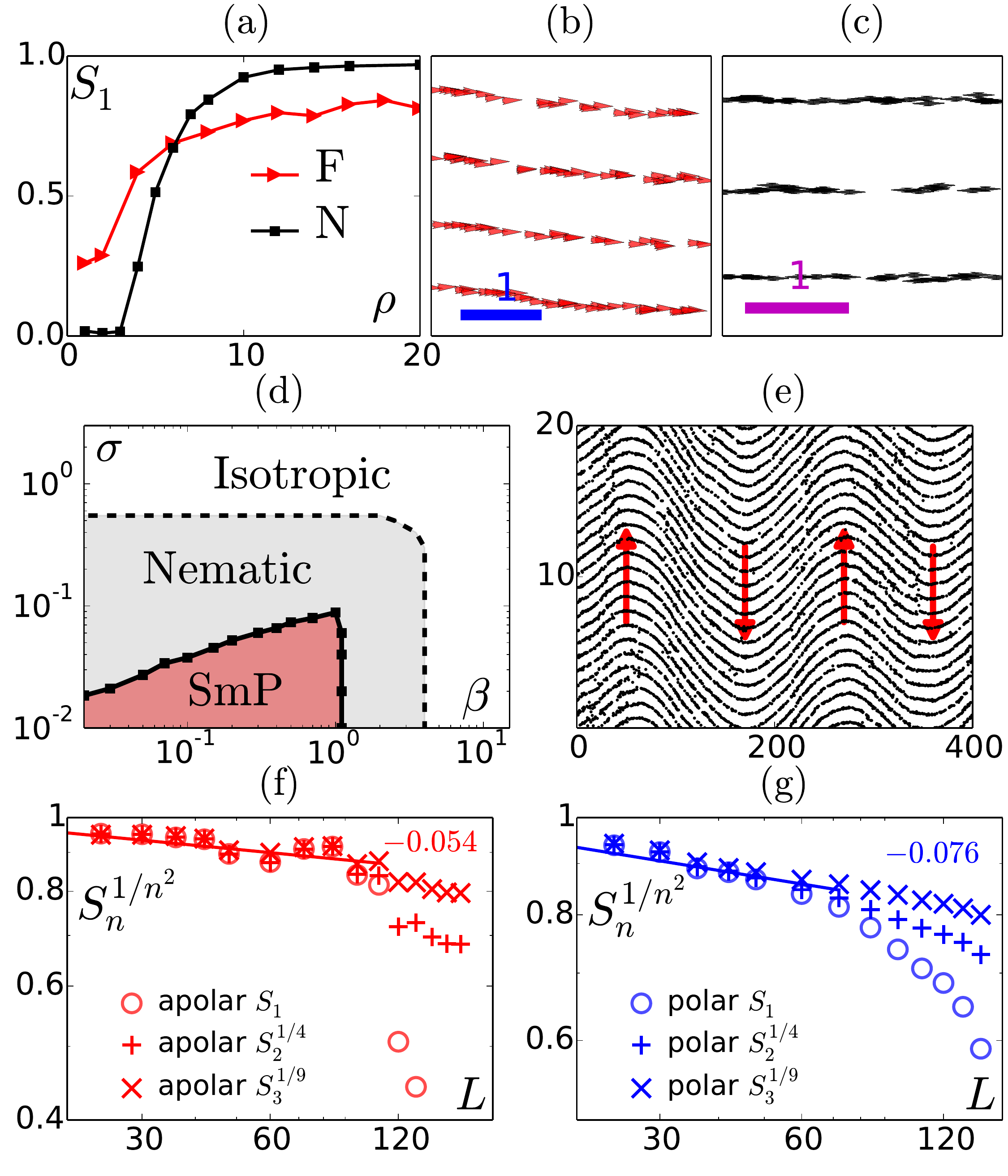}
\end{center}
\caption{Simulation results for isotropic repulsion. 
(a) 
Smectic order parameter $S_1$ as a function of global density. Black squares : apolar model with $\beta=0.2$, $\sigma=0.02$; red triangles: polar model with  $\beta=0.32$, $\sigma=0.04$.
(b,c) Smectic-P configurations for the polar model with $\rho_0=10$, $\beta=0.05$, $\sigma=0.007$ and the apolar model with  $\rho_0=8$, $\beta=0.1$, $\sigma=0.01$. Only a small portion of a bigger system  is shown. 
(d) Phase diagram in the $(\sigma,\beta)$ plane for the apolar model with $\rho_0=8$.
(e) Undulations for a system of size $L=400$ in the apolar model {$\rho_0=8$, $\beta=0.1$, $\sigma=0.01$}. Part of a configuration with the $y$ direction dilated for clarity. Arrows indicate the slow motion of the arches.
(f,g) Scaling of smectic order parameters $S_n$ with system size $L$. (f) The apolar model with   $\rho_0=8$, $\beta=0.1$, $\sigma=0.01$ shows a weak algebraic decay of $S_1$ (red circles) up to a critical system size $L_c\approx100$ with $S_2^{1/4}$ (red crosses) and $S_3^{1/9}$ (red oblique crosses) falling perfectly on top of $S_1$ for $L\ll L_c,$ as predicted by the hydrodynamic theory. Above $L_c$ scaling breaks down due to the undulation instability.
(g) The polar model with  $\rho_0=10$, $\beta=0.05$, $\sigma=0.007$  shows similar weak algebraic decay at small $L\ll L_{NL}\approx60$. However for $L\gg L_{N\!L}$ the results for $S_1$, $S_2^{1/4}$, and $S_3^{1/9}$ start to diverge, and $S_1$ decays faster than algebraically. 
\label{Fig1-IsoRep}
}
\end{figure}
%%%%%%%%%%%%%%%%%%%%%%%%%%%%%%%%%%%%%%%%%%%%%%%%%%%%%%%%%%%%%%%%%%%%%%%

%%%%%%%%%%%%%%%%%%%%%%%%%%%%%%%%%%%%%%%%%%%%%%%%%%%%%%%%%%%%%%%%%
\subsection{Regions of smectic order}
%%%%%%%%%%%%%%%%%%%%%%%%%%%%%%%%%%%%%%%%%%%%%%%%%%%%%%%%%%%%%%%%%

For $\beta=0$, the polar  and apolar models defined above respectively  reduce exactly to the Vicsek model \cite{vicsek} and the active nematics 
model of \cite{chatenematicPRL2006},  which both have two main parameters: the noise strength $\sigma$
and the global mean number density $\rho_0$. Thus, $\beta$, which controls the 
repulsion strength, is the only new parameter. 
We have explored systematically the three-parameter space of our models. 
As in the repulsion-free case,
 an {\it orientational} order-disorder transition is found upon increasing $\sigma$ and/or decreasing
$\rho_0$, in both the polar and the apolar case. {Note that the coexistence phase of this transition, best described in the 
liquid/gas framework \cite{phase-sep}, is essentially unobservable at such high-densities.}
Significant smectic order is only found for relatively high densities ($\rho_0\gtrsim 8$), 
inside the orientationally-ordered phase (Fig.~\ref{Fig1-IsoRep}a). 

We quantify this smectic order with smectic order parameters $S_n$, which are defined as
\begin{eqnarray}
S_n\equiv I({\bf q}_n,t) \,,
\end{eqnarray}
where $I({\bf q},t)\equiv{\langle|\rho({\bf q},t)|^2\rangle/ N^2}$, 
with $N=\rho_0 L^2$ the total number of particles, and ${\bf q}_n\equiv nq_0 \hat{z}$, with $q_0\equiv{2\pi\over a_l}$. 
%\sout{the mean squared amplitudes of the  Fourier transformed density at wavevectors ${\bf q}=n{2\pi\over a}\bf{\hat{z}}$,} 
Here
\begin{eqnarray}
 \rho ({\bf q},t)=\int d^2r \, \rho({\bf r},t)e^{-i{\bf q}\cdot{\bf r}}\,.
\end{eqnarray}
is the spatial Fourier transform of the density,
$n$ is a non-zero integer, $\bf{\hat{z}}$ is the mean normal to the layers, and $a_l$ the layer spacing,
which is typically  of the order of the interaction range, {see  \ref{App:smop}.
We hereafter restrict ourselves to high average densities ($\rho_0\geq 8$) and low speeds ($v_0=0.2-0.3$), and vary $\sigma$ and $\beta$. 
We use square domains of linear size $L$ with periodic boundary conditions, and consider only 
``zero-winding number" smectic configurations (see  \ref{App:wind} for details).

The polar and apolar models, whether with isotropic or anisotropic repulsion, 
share qualitatively similar phase diagrams in the ($\beta$,$\sigma$) plane, 
whose main features are shown in Fig. \ref{Fig1-IsoRep}d for the apolar model with isotropic repulsion. 
When both repulsion and noise strength $\sigma$  are sufficiently low, the system develops orientational order.
Inside this region, a line starting at $\beta=\sigma=0$ and finishing at $\beta_{\rm max}$ close to the orientational order-disorder transition delimits a domain in which  smectic order appears over a large, but finite, range of length scales (cf.  \ref{App:smop}, Fig. \ref{figApp}). 
The location of the boundaries of these regimes do not change much with system size $L$, but
the maximal levels of {\it global} smectic order generally decrease with increasing $L$ (whereas {\it local} smectic order remains strong).

The type of smectic order observed depends on the model (Table~\ref{table}).
For isotropic and anisotropic repulsion with $\gamma=\frac{\pi}{2}$, a novel type of smectic 
order  emerges
that is {\it not} observed in equilibrium \cite{deGennes}.  
We call this new type of smectic order, in which the particle axes are {\it parallel} to the layers (Figs.~\ref{Fig1-IsoRep}b,c \& \ref{Fig3-SmPict}b,d), ``smectic P".

Since repulsion between particles favors translational order, {\it anisotropic} repulsion favors {\it anisotropic} translational order (i.e., layering).  This implies, in particular, that when  repulsion is stronger perpendicular to the direction of particle motion (as is the case for case $\gamma=\pi/2$), the smectic P configuration just described is preferred. 

In our case, isotropic repulsion leads to the smectic P phase, which also occurs, for the reasons just given, in the explicitly anisotropic case $\gamma=\pi/2$, for which  repulsion is stronger to the left and right of the direction of motion of the particles. This is because, in the isotropic case, particles do not move away from those ahead of or behind them as effectively, since this requires speeding up or slowing down, which in the simplest model with fixed speed they cannot do. We have confirmed that the P phase is also favored for active particles models with variable speed \cite{ABP}, in which the speeds of the particles are allowed to fluctuate around the preferred speed $v_0$ due to the action of the various forces. In general, the additional anisotropy linked to the propulsion direction of self-propelled particles effectively mimics the effect of weaker repulsion in those directions.

For $\gamma=0$, repulsion is strongest in the front and back of particles, and 
one might naively expect smectic-A order. 
However, in the apolar case, this is in fact never observed. Instead, the left/right symmetry is 
spontaneously broken, and the particles' axes tip away from
the layer's normal: a smectic C with $|\Phi-\Phi_{\rm S}| \simeq 0.2\pi$, 
where $\Phi$ and $\Phi_{\rm S}$ denote respectively the angles of the particle axes and the 
layer normal (cf. Fig \ref{Fig2-AniRep} d). 
Only in the polar model, after a possibly long transient during which patches of smectic-C order compete and form locally chevron-like structures, Fig.~\ref{Fig2-AniRep}b), does the system eventually reach a smectic-A state (Fig.~\ref{Fig2-AniRep}c).

%%%%%%%%%%%%%%%%%%%%%% TABLE %%%%%%%%%%%%%%%%%%%%%%%%%%%%%%%%%%%%%%%%%
\begin{table}[tbp!]
	\begin{center}
\caption{Type of smectic phase (A, C, P) and large-scale phenomena observed 
for polar and apolar models with isotropic and anisotropic ($\gamma=0, \frac{\pi}{2}$) repulsion.
\label{table}
}
\begin{tabular}{llll}
	repulsion: & isotropic & $\gamma=\frac{\pi}{2}$ & $\gamma=0$ \\ \hline
polar model: & P, travelling fluctuations & P, rotation & A, swirls \\
apolar model: & P, undulations & P, undulations & C
\end{tabular}
\end{center}
\end{table}
%%%%%%%%%%%%%%%%%%%%%%%%%%%%%%%%%%%%%%%%%%%%%%%%%%%%%%%%%%%%%%%%%%%

%%%%%%%%%%%%%%%%%%%%%%%%%%%%%%%%%%%%%%%%%%%%%%%%%%%%%%%%%%%%%%%%%%%%%%
\subsection{Quasi-long-range order and large-scale phenomena}
%%%%%%%%%%%%%%%%%%%%%%%%%%%%%%%%%%%%%%%%%%%%%%%%%%%%%%%%%%%%%%%%%%%%%%

In all cases, defect-less smectic configurations are only observed in sufficiently small (albeit still quite large) systems (see Supp. Mat. Movies 1 \& 2). 
Even in simulations that start from carefully prepared ``perfect" initial configurations, spontaneous nucleation of dislocations
and/or large-scale undulations occur beyond some model-dependent lengthscale, which we call $L_{\rm c}$ ($L_{N\!L}$) 
for the apolar (polar) case; the meaning of these names is explained in the ``Hydrodynamic theory" section below.
For $L$ less than this length scale, $S_n$ shows either no decay,  or very slow algebraic decay with 
$L$; the latter is what we mean by quasi-long-range smectic order.
For much larger systems, $S_n$ decreases much faster with $L$ (Fig.~\ref{Fig1-IsoRep}f,g).
For $L\gg L_{\rm c}$ in the apolar case, the system shows undulations of the smectic-P layers (Fig.~\ref{Fig1-IsoRep}e). 
The wavelength of the undulations is well defined asymptotically and typically large ($\sim100-200$). 
Note that this wavelength is of order of $L_{\rm c}$. 
This undulation pattern is not quite steady: its ``arches" slowly move as indicated by the arrows in Fig.~\ref{Fig1-IsoRep}e and Supp. Mat. Movie 3. This accompanied by the constant nucleation of dislocations in the sheared regions between arches.
For $L\gg L_{N\!L}$ in the polar case, no undulation instability is observed. At large scales, polar smectics-P layers show traveling fluctuations on all scales. In contrast to the apolar case, these fluctuations never cohere into a regular moving pattern that is stationary in a co-moving frame. Instead, the deviations continue evolving randomly in time, and  do not develop a single typical wavelength. Note that our theory predicts that fluctuating modes with the largest wavelength are expected to have the largest amplitude, which may produce the impression of a transient regular pattern with wavelength of the order $L/2$ (Supp. Mat. Movie 2).

Conspicuous by their absence in our simulations of both apolar and polar systems are the ``giant number fluctuations'' \cite{tonertuPRL1995, sradititonerEPL2003,GNF, Narayanan-jstat-2006} 
(cf. Fig. \ref{Fig4-SmGNF})  observed \cite{GNF,Narayanan-jstat-2006}, and predicted by hydrodynamic theories \cite{tonertuPRL1995, GNF} in both polar \cite{tonertuPRL1995, GNF} and 
apolar \cite{sradititonerEPL2003, Narayanan-jstat-2006} flocks with orientational, but no translational, order.

%%%%%%%%%%%%%%%%%%%%%%%%%%%%%% FIGURE 2 %%%%%%%%%%%%%%%%%%%%%%%%%%%%%%%
\begin{figure}[tbph]
	\begin{center}
 		{\includegraphics[width=0.8\textwidth]{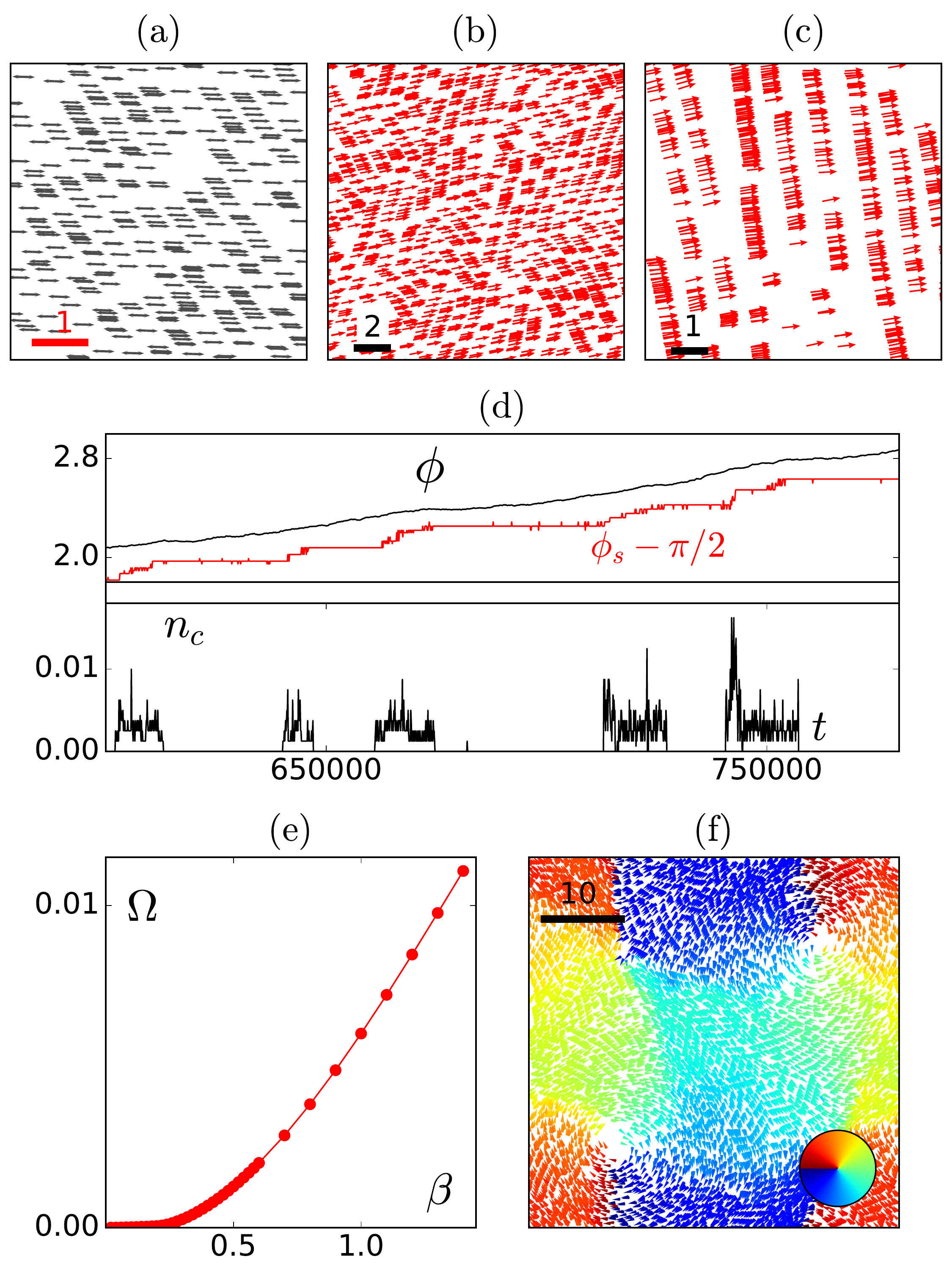}}
\caption{Anisotropic repulsion.
	(a,b,c) Configurations with $\gamma=0$ (scale indicated by the bar). 
(a) Apolar model with a portion of a smectic-C pattern ($L=40$, $\rho_0=20$, $\sigma=0.005$, $\beta=0.1$)
(b) Transient pattern with chevron structures in the polar model, due to competition of differently tilted smectic-C patches.  ($L=24$, $\rho_0=10$, $\sigma=0.008$, $\beta=0.128$).
(c) Portion of a final smectic-A configuration  ($L=20$, $\rho_0=12$, $\sigma=0.05$, $\beta=0.5$).
(d,e) Rotation of polar, smectic P ($\gamma=\frac{\pi}{2}$, $L=40$, $\rho_0=10$, $\sigma=0.02$). (d) Time series for $\beta=0.04$ ; top panel: orientation of polar ($\phi$) and smectic ($\phi_s$)  directors, the latter angle being shifted by $-\pi/2$ ; bottom panel: density of defects. 
(e) Rotation speed $\Omega$ as a function of $\beta$.
%(f) Ordinary number fluctuations in the presence of local smectic-P order in the apolar model with isotropic repulsion up to system sizes beyond $L_c$ ($\rho=8$, $v=0.3$, $\sigma=0.01$, $\beta=0.1$). 
(f) Swirls of polar particles with anisotropic repulsion ($\gamma=0$); color indicates direction of motion ($L=48$, $\rho_0=10$, $\sigma=0.01$, $\beta=1.3$). 
\label{Fig2-AniRep}
}
\end{center}
\end{figure}
%%%%%%%%%%%%%%%%%%%%%%%%%%%%%%%%%%%%%%%%%%%%%%%%%%%%%%%%%%%%%%%%%%%% 
%%%%%%%%%%%%%%%%%%%%%%%%%%%%%% FIGURE 3 %%%%%%%%%%%%%%%%%%%%%%%%%%%%%%%
\begin{figure}[tbp!]
	\begin{center}
 	\includegraphics[width=0.65\textwidth]{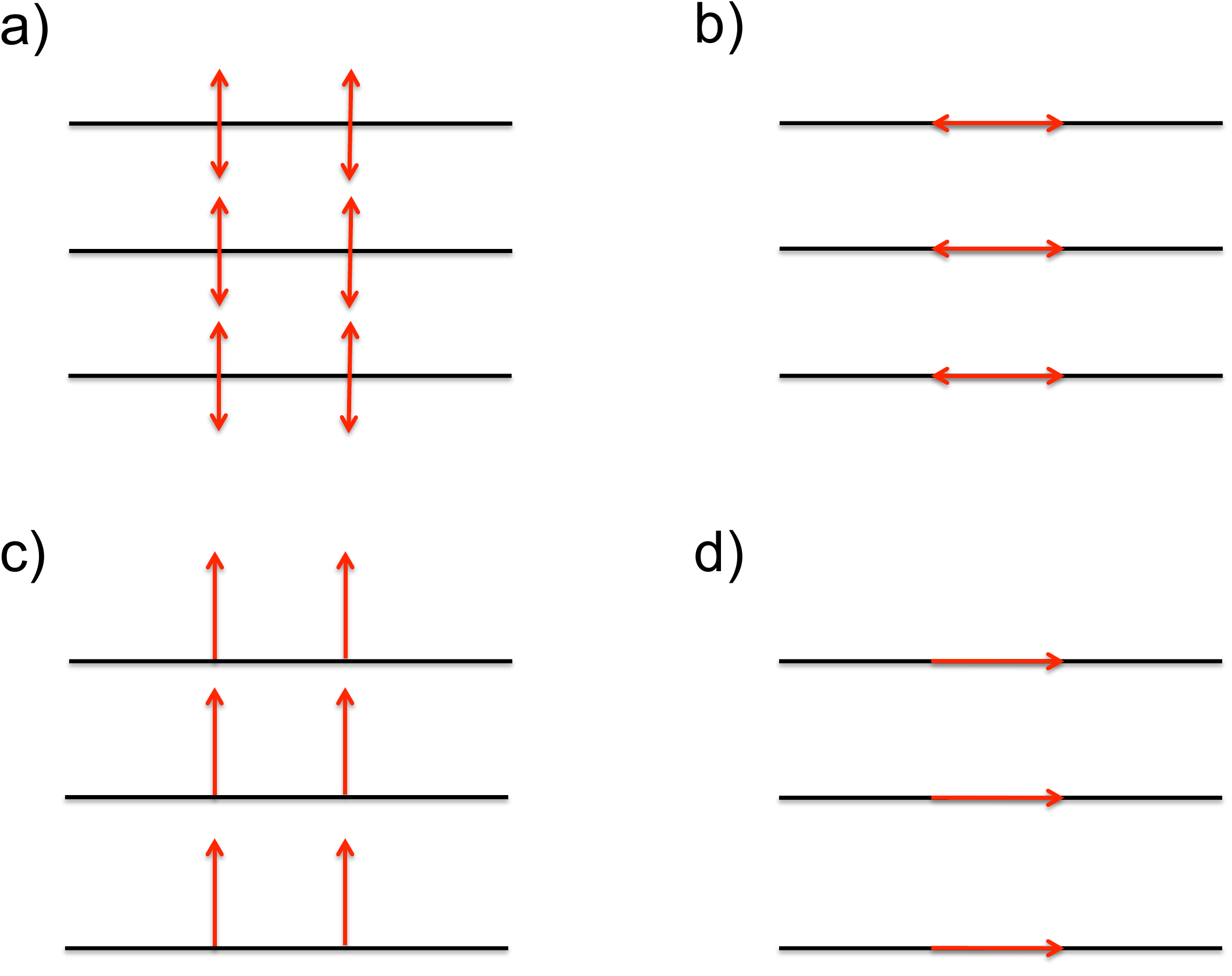}
\end{center}
\caption{Pictorial illustration of the different types of smectic order. Red arrows indicate the dominant direction(s) of particle motion in each case.
(a) Apolar Smectic A. (b) Apolar smectic P. Note that this has the same symmetries (in particular, both up-down and left-right inversion symmetry) as the apolar A phase.  (c) Polar smectic A phase. This has lost up-down symmetry, but retains left-right symmetry.
(d) Polar smectic P. This has {\it lost} left-right symmetry, but {\it retains} up-down symmetry. It therefore is a distinct phase from the polar A phase. 
\label{Fig3-SmPict}
}
\end{figure}
%%%%%%%%%%%%%%%%%%%%%%%%%%%%%%%%%%%%%%%%%%%%%%%%%%%%%%%%%%%%%%%%%%%%

We finally report on two spectacular phenomena observed with anisotropic repulsion.
For polar smectics P with $\gamma=\frac{\pi}{2}$,
we observe a spontaneous breaking of the left/right symmetry in the form of global rotation of both 
the particle axes and the smectic layers. 
For small $\beta$ values, at fixed system size, rotation may be intermittent and may change sign
in time (Fig.~\ref{Fig2-AniRep}d and Supp. Mat. Movie 4).
But for sufficiently large systems, even an initially prepared defect-free smectic-P configuration
starts rotating either clockwise or counterclockwise with a well-defined, steady 
angular velocity $\omega$ which increases strongly with $\beta$ (Fig.~\ref{Fig2-AniRep}e).
Our observations of the intermittent, slow-rotation regimes reveal that the polar orientation angle
$\Phi$ is `driving' rotation: the angle of particle axes changes first, with $\Phi_{\rm S}$, the angle of the layer's normal,  fixed, lagging behind. The stress thus induced on the smectic layers then generates dislocations, $\Phi_{\rm S}$ slips, and the defects annihilate. Since  $\Phi$ continues rotating, this sequence continuously  repeats itself. We note that we do not observe hopping of individual particles between the layers;  instead the observed global rotation corresponds to breaking of layers and reattaching to neighboring layers, which leads to formation of dislocation lines travelling through the system ("defect waves", Supp. Mat. Movie 4).

For the polar smectic-A, another spectacular phenomenon sometimes occurs
for the larger $\beta$ values within the smectic region: large-scale rotating ``swirls'', made of a spiral, flower-like arrangement of smectic layers (Fig.~\ref{Fig2-AniRep}g, and Supp. Mat. Movie 5). In a periodic domain, a number of clockwise and counterclockwise swirls may emerge, which 
experience some effective repulsion, leading to stable configurations of equal number of `$+$' and `$-$' swirls that coexist with the swirl-free one.

%%%%%%%%%%%%%%%%%%%%%%%%%%%%%% FIGURE 4 %%%%%%%%%%%%%%%%%%%%%%%%%%%%%%%
\begin{figure}[t]
	\begin{center}
 		{\includegraphics[width=0.6\textwidth]{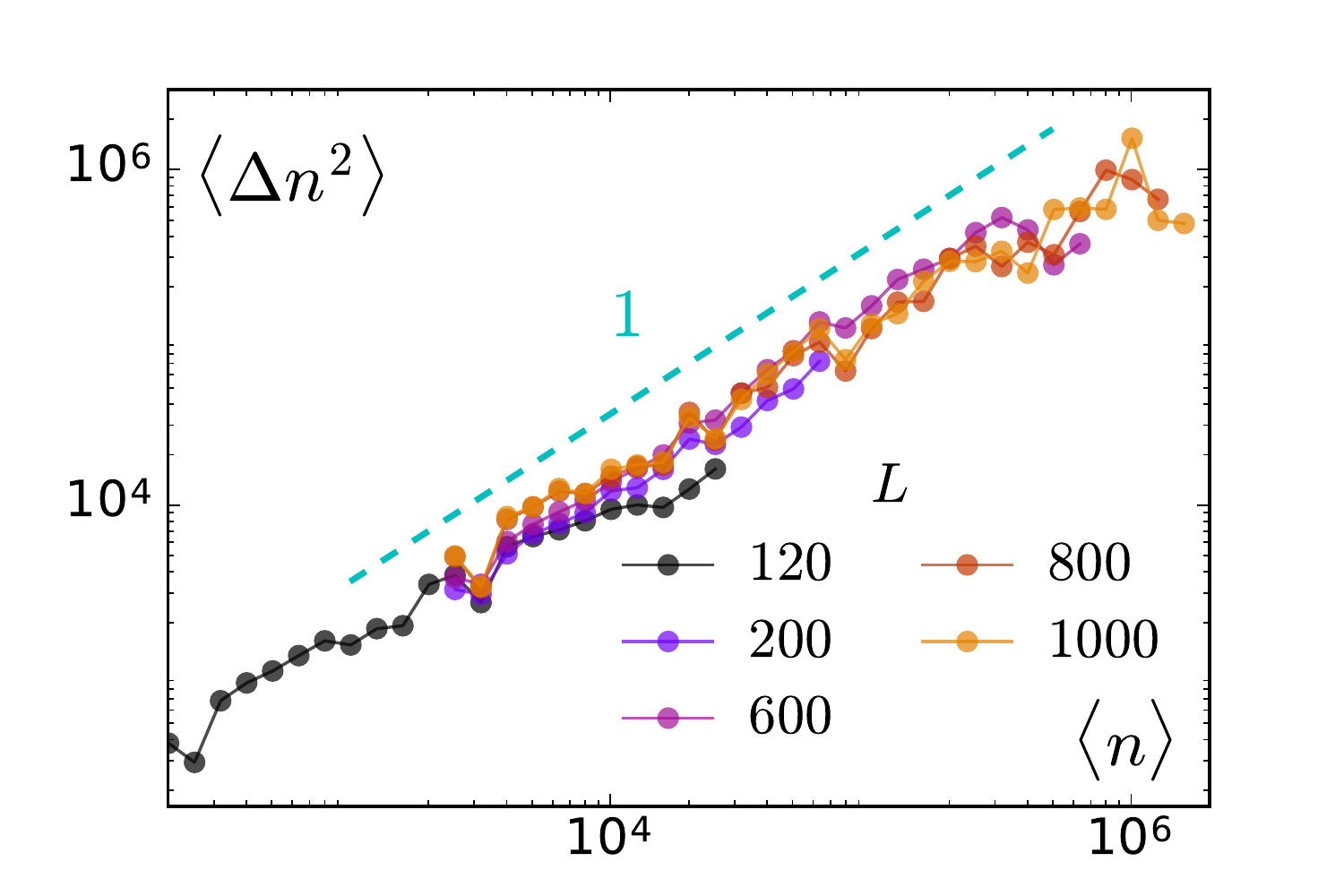}}
\caption{
	Number fluctuations  for apolar particles with nematic interactions and isotropic repulsion. These systems exhibit  smectic-P order for system sizes $L\ll L_c$, and instability for $L\gtrsim L_c$. The critical system size for the parameters used here is $L_c\approx200$. 
	In contrast to both polar and apolar active fluids with orientational, but no translational, order, we observe no ``giant number fluctuations'': that is, we do {\it not} observe $\langle\Delta n^2\rangle \propto \langle\Delta n^2\rangle^\alpha$ with $\alpha>1$; rather, we find conventional behavior (i.e., $\alpha=1$).
Parameters: $\rho=8$, $v=0.3$, $\sigma=0.01$, $\beta=0.1$
\label{Fig4-SmGNF}
}
\end{center}
\end{figure}
%%%%%%%%%%%%%%%%%%%%%%%%%%%%%%%%%%%%%%%%%%%%%%%%%%%%%%%%%%%%%%%%%%%%

%%%%%%%%%%%%%%%%%%%%%%%%%%%%%%%%%%%%%%%%%%%%%%%%%%%%%%%%%%%%%%%%%%%%%
\section{Hydrodynamic theory}
%%%%%%%%%%%%%%%%%%%%%%%%%%%%%%%%%%%%%%%%%%%%%%%%%%%%%%%%%%%%%%%%%%%%%

We now demonstrate that most of the above results are well accounted for by hydrodynamic theories of active smectics. 
The variables of the hydrodynamic theory are the  
field $u({\bf r}, t)$ giving the displacement of the layers  perpendicular to their unperturbed, %\sout{from some reference set of } 
uniformly spaced, parallel %\sout{layers} 
locations, and the number density $\rho({\bf r}, t)$. The latter is a slow hydrodynamic variable because of number conservation, while the former is the Goldstone mode associated with the spontaneous breaking of continuous translational symmetry in the direction perpendicular to the layers.
The theory consists of a closed set of stochastic partial differential equations for the time evolution of $u(\bf{r},t)$ and $\rho(\bf{r},t)$, containing all terms  at lowest order in a gradient expansion that are consistent with the symmetries of the underlying dynamics and the broken symmetry state. 
Since we have two different broken symmetries (polar and apolar smectic P), we have two sets of hydrodynamic equations.

%%%%%%%%%%%%%%%%%%%%%%%%%%%%%%%%%%%%%%%%%%%%%%%%%%%%%%%%%%%%%%%%%%%%%
\subsection{Apolar smectic P}
%%%%%%%%%%%%%%%%%%%%%%%%%%%%%%%%%%%%%%%%%%%%%%%%%%%%%%%%%%%%%%%%%%%%%
In two dimensions, there is no symmetry difference between the smectic-P order found here 
and the apolar active smectic-A phase treated in \cite{live-soap}: both are symmetric under separate inversions about the  $z$ axis (the normal to the smectic layers) and about the $x$ axis (along the layers). Hence the equations for this case are the same as those developed in \cite{live-soap} for apolar active smectic A. They read
\begin{eqnarray}
\partial_{t} u &\!\!\!=& B\partial_{z}^2 u + D_{ux}\partial_x^2 u - K  \partial_x^4 u+ C\partial_z\delta\rho+f_u 
\label{EOM_u}\\
 \partial_{t} \delta\rho &\!=\!& D_{\rho x}\partial_x^2\delta\rho + D_{\rho z}\partial_z^2\delta\rho
 +C_ x\partial_z\partial_x^2u+ C_{z}\partial_z^3u +f_\rho
\label{EOM_rho}
\end{eqnarray}
where $\delta\rho\equiv\rho-\rho_0$, and $f_u$ and  $f_\rho$ are Gaussian, zero-mean, white noises with variances 
\begin{eqnarray}
\langle f_u({\bf r}, t)f_u({\bf r}', t')\rangle &\!\!\!=\!\!\!& \Delta_u\delta({\bf r} \!-\! {\bf r}')\delta(t \!-\! t')\label{fucor} \\
\langle f_\rho({\bf r}, t) f_\rho({\bf r}', t')\rangle &\!\!\!=\!\!\!& (\Delta_{\rho x} \partial_{ x}^2+\Delta_{\rho z} \partial_z^2) \delta({\bf r} \!-\! {\bf r}') \delta(t \!-\!t')
\label{frhocor}
\end{eqnarray}
We set the cross-correlation $\langle f_\rho({\bf r}, t) f_u({\bf r}', t')\rangle=0$; corrections to this can be shown to be ``irrelevant" in the renormalization group sense of having no effect on the large distance, long time behavior of the system.
Here $B$, $D_{ux}$,  $K$, $D_{\rho x}$, $D_{\rho z}$, $C$, $C_x$, $C_z$, $\Delta_u$,  $\Delta_{\rho x}$, 
and $\Delta_{\rho z}$ are all phenomenological parameters that cannot be determined by symmetry 
arguments, but must be  deduced from experiments, simulations or a detailed kinetic theory.

In an equilibrium smectic, %without momentum and number conservation,
the 
$D_{u x}$ term in (\ref{EOM_u}) is forbidden by rotation-invariance of the free energy. It is, however,
permitted here \cite{active tension1} simply because rotation-invariance \textit{at the level of the
EOM}, which is all one can demand in an active system, does not
rule them out. The physical content of this, term is that layer curvature produces a local vectorial asymmetry which must lead to directed motion of the layers, as this is a driven system. A similar term occurs in single membranes with ``pumps" \cite{active tension1, active tension2}.

The $B$ term simply acts to keep the smectic layers equidistant, while the $K$ term is a bend modulus that tends to keep them straight. The rest of the terms in (\ref{EOM_u}) and  (\ref{EOM_rho}) are present in equilibrium, although the equilibrium requirement \cite{MRT} that ${C_x\over C_z}={D_{\rho x}\over D_{\rho z}}$ does not hold in our non-equilibrium system.

%\sout{Note that equations for an {\it equilibrium} smectic without momentum conservation are a special case of  Eqs.~\eref{EOM_u} and \eref{EOM_rho}, with the constraint ${C_x\over C_z}={D_{\rho x}\over D_{\rho z}}$ and, most importantly, the ``active tension" $D_{ux}=0$. }

There is no {\it a priori} symmetry argument that determines the sign of $D_{u x}$.
As shown in \cite{live-soap}  when $D_{u x}>0$, smectic layer fluctuations are suppressed, 
leading to quasi-long-range order in $d=2$, in contrast to the short-ranged order found in equilibrium \cite{1st}.
When $D_{u x}<0$, the perfect smectic state with constant $u({\bf r}, t)$ and   
$\rho({\bf r}, t)=\rho_0$ is unstable against ``undulations",  in which the layers wiggle in unison.
This is easily seen by Fourier transforming the (linear) equations of motion in space.
For wavevectors $\bf{q}$ with $q_z=0$, $\rho$ and $u$ decouple, leading to the growth rate %dispersion relation 
$\nu(q_x)=- D_{ux}q_x^2  - K  q_x^4$. 
Assuming $K>0$, all modes with $q_x<q_{\rm max}\equiv\sqrt{|D_{u  x}|/K}$ are unstable, with the most unstable mode
being $q_{\rm c}=q_{\rm max}/\sqrt{2}$. We show in { \ref{eig_inst}} that allowing $q_z\ne 0$ does not change this result: 
the instability always grows fastest along $x$.
This is precisely what our simulations show: 
undulations appear and grow along the layers, but only in systems whose extension $L_x$ along this axis is large enough. 
This minimal size is of the order of the scale of the observed undulations, which we identify with:
\begin{eqnarray}
	L_{\rm c}\equiv\frac{2\pi}{q_{\rm max}}  =2\pi\sqrt{K \over |D_{u  x}|}\,.
\label{Lcdef}
\end{eqnarray}  
Note that the wavelength of the undulations is of order $L_{\rm c}$, but not necessarily equal to it, as it will 
ultimately be determined by nonlinear, saturating terms neglected here.
We will also use Eq.~\eref{Lcdef} to define a ``critical length scale" in the {\it stable} case $D_{ux}>0$, 
as the scale beyond which active tension effects (which then {\it suppress} fluctuations) become important.

We now analyze noise-induced fluctuations in our theory. We focus on the case in which
$L_{\rm c}$ is large, the positional noise $\Delta_u$ is small, and the density noises $\Delta_{\rho (x,z)}$ 
are large, all of which, as we argue in {\ref{App:UuApolar}}, are satisfied in our simulations. 
Here we simply sketch the derivation of, and summarize, our results; details are given in {Appendices A.3 and A.4}.

We begin by spatiotemporally Fourier transforming the equations of motion \eref{EOM_u} and \eref{EOM_rho} and solve the resultant linear algebraic equations for the transformed fields $u(\bf{q},\omega)$ and $\rho(\bf{q},\omega)$ in terms of the random forces $f_u(\bf{q},\omega)$ and $f_\rho(\bf{q},\omega)$. 
Autocorrelating, and using \eref{fucor} and \eref{frhocor} gives an expression for $C_{uu}({\bf q}, \omega)\equiv\langle|u({\bf q}, \omega)|^2\rangle$. Integrating this over all frequencies $\omega$ gives
the equal time correlation function $C^{ET}_{uu}({\bf q})\equiv\langle|u({\bf q},t)|^2\rangle$, 
from which we can in turn calculate equal-time real space correlation functions. 

%%%%%%%%%%%%%%%%%%% FIGURE 3 %%%%%%%%%%%%%%%%%%%%%%%%%%%%%%%%%%%%%%%%%%
\begin{figure}[tb]
\begin{center}
 \includegraphics[width=0.7\textwidth]{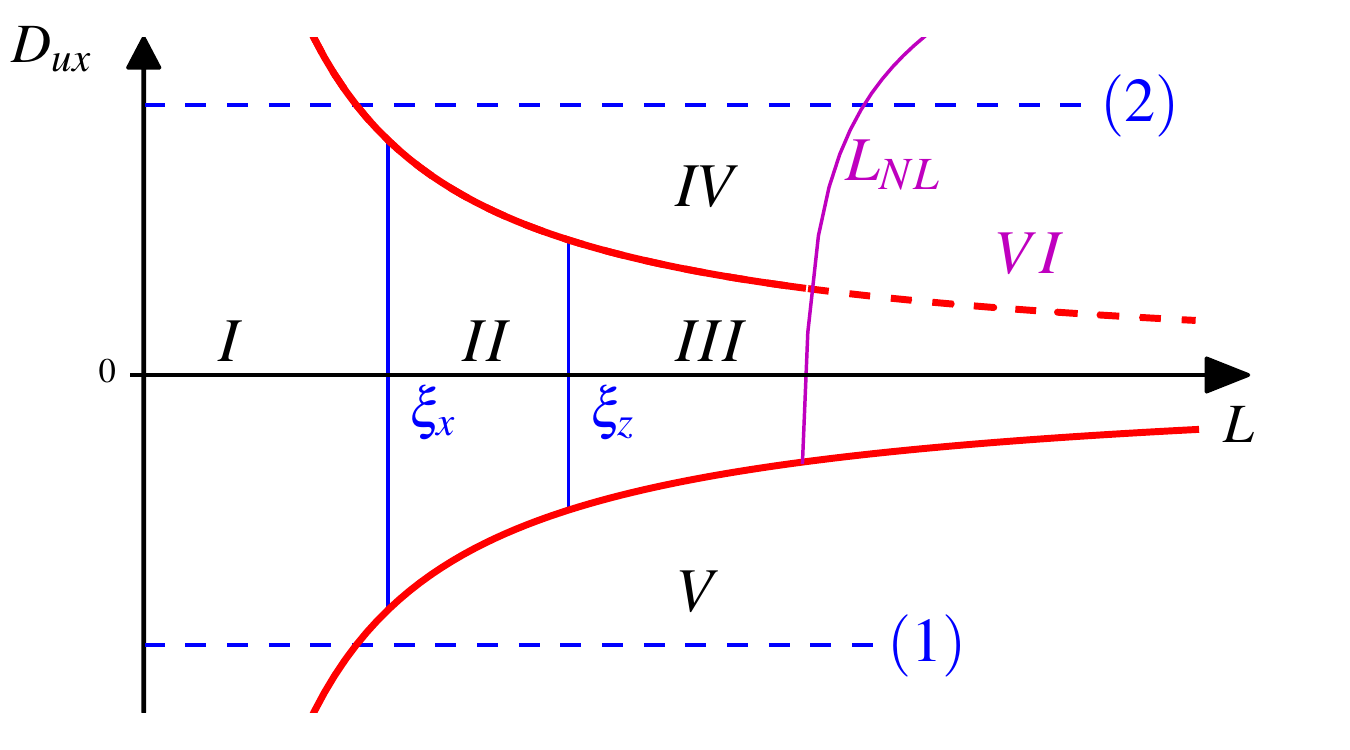}
\end{center}
\caption{Regions distinguished by the scaling behavior of smectic order parameters $S_n$ with system size $L$ and active tension $D_{ux}$. 
Red lines indicate the lengthscale $L_{\rm c}$ defined in the main text.
In region I, defined by $L< L_{\rm c}\equiv
2\pi\sqrt{K/|D_{u x}|}$ and $L \ll \xi_x$, in the apolar case, 
$S_n\propto L^{-n^2\eta_I}$; in the polar case, $S_n$'s do not decay. 
In region II, defined by $L<L_{\rm c}$ 
and $\xi_x\ll L \ll \xi_z$, both the apolar and polar smectic P obeys $S_n\propto  L^{-1-\eta_In^2}n^{-2}$.
In region III ($L < L_{\rm c}$ and $ L \gg \xi_z$), both the apolar and polar smectic P obey $S_n\propto L^{-2}n^{-6}$: 
purely short-ranged smectic correlations, as explained in the main text. 
In region IV ($L\gg L_{\rm c}$ and $D_{u x}>0$) $S_n\propto L^{-n^2\eta_{IV}}$ for {\sl both} the apolar and polar cases. 
In region V, which is $L\gg L_{\rm c}$ and $D_{ux}<0$, both the apolar and polar smectics P are unstable, 
and we expect $S_n\propto L^{-2}$. 
Region VI, in which $L\gg L_{N\!L}$,  only exists for the polar case. Here nonlinear effects become important, and may induce dislocations 
that destroy the smectic order. If they do, then we expect $S_n\propto L^{-2}$ in this region as well.  
The horizontal dashed lines above and below the $L$-axis are the path followed by increasing $L$ at fixed parameters in our simulations of respectively 
the apolar model (1, data shown in Fig.~\ref{Fig1-IsoRep}f) and the polar model (2, data shown in Fig.~\ref{Fig1-IsoRep}g).
\label{Fig5-TheoR}
}
\end{figure}
%%%%%%%%%%%%%%%%%%%%%%%%%%%%%%%%%%%%%%%%%%%%%%%%%%%%%%%%%%%%%

The correlations of $u$ we thereby obtain are related to the smectic order parameters.
Recall that $S_n\equiv I({\bf q}_n,t)$, where $I({\bf q},t)\equiv{\langle|\rho({\bf q},t)|^2\rangle/ N^2}$, 
with $N=\rho_0 L^2$ the total number of particles, and ${\bf q}_n\equiv nq_0 \hat{z}$. 
These $S_n$ are sensitive to fluctuations of the displacement field $u$ since a translation by $\Delta u$ 
clearly changes the complex phase of $\rho({\bf q}_n,t)$ by $nq_0\Delta u$.
%%%HC changed u into \Delta u above 
We show in \ref{smop_apolar} that for the apolar model:
\begin{eqnarray}
S_n=\frac{w_n}{L^2} \int_0^Ldx\int_0^Ldz \exp\left[-{n^2q_0^2\over 2}\langle|\Delta u({\bf  r})|^2\rangle\right]
\label{snscale1}
\end{eqnarray}
where $\langle|\Delta u({\bf r})|^2\rangle\equiv\langle|u({\bf r}+{\bf r}',t)-u({\bf r}\,',t)|^2\rangle$
can be easily obtained by Fourier transforming %\eref{uqt}  
$C^{ET}_{uu}({\bf q})$
back to real space, and the $w_n$'s are uninteresting constants given in { Eq. \eref{Wn}}.
This calculation, given in detail  in {\ref{real_uu}}, reveals the existence of two 
other important lengths in addition to $L_c$:
\begin{eqnarray}
	\xi_x\equiv{4\over q_0}\sqrt{{3K\over \Delta_u}}, \quad \textnormal{and} \quad 
\xi_z\equiv{16\pi\over q_0^2}{\sqrt{KB'^3}\over \Delta_u^2},
\end{eqnarray}
with $B'\!\equiv\! B\!-\!CC_x/D_{\rho x}$.
The resultant behavior of 
$S_n$ with system size exhibits numerous crossovers, which delimit five different regions of the 
$(L,D_{ux})$ plane, as  summarized in Fig.~\ref{Fig5-TheoR}:

\begin{description}
	\item[Region I] is defined by $L\ll L_{\rm c}\equiv
\sqrt{K / |D_{u x}|}$ and $L \ll \xi_x$.  Here the order parameters $S_n$ decay 
algebraically with $L$: $S_n\propto L^{-n^2\eta_{_{_I}}}$, with 
$ \eta_{_{_I}}$ a {\it non-universal} exponent 
that is a monotonically increasing function of the noise strengths 
$\Delta_{\rho x}$ and $\Delta_{\rho z}$.  
	\item[Region II,] in which $L\ll L_{\rm c}$ and $\xi_x\ll L \ll 
\xi_z$, is characterized by rather fast algebraic decay of $S_n$ according to  $S_n\propto  L^{-1-\eta_{_{_{I}}} n^2}n^{-2}$.
\item[Region III,] defined by $L \ll L_{\rm c}$ and $ L \gg \xi_z$, shows even faster decay with $L$: $S_n\propto 
L^{-2}n^{-6}$.  
This corresponds to purely short-ranged smectic correlations. 
the system breaks up into decorrelated smectic regions of fixed size whose number scales as $L^2$; 
their contributions to the global order parameters $S_n$ simply add randomly, leading to the scaling $S_n\propto{1\over N_r}\propto{1\over L^2}$.
\item[Region IV,] in which $L\gg L_{\rm c}$ and $D_{u x}>0$, exhibits quasi long-ranged smectic order 
with $S_n\propto L^{-n^2 \eta_{_{_{I\!V}}}}$, where $\eta_{_{_{I\!V}}}$ is very close to, but different from 
$\eta_{_{_{I}}}$ (see  Eq. \ref{eta4})
\item[Region V] is defined by $L\gg L_{\rm c}$ and $D_{u x}<0$. Here, the system is unstable against undulations, and we expect $S_n \propto L^{-2}$.
\end{description}

Note that varying $L$, while keeping parameters fixed, will not necessarily explore all of these regions. 
For example, a system with a sufficiently large in magnitude, negative $D_{u x}$ (specifically, large 
enough that $L_{\rm c}
\ll\xi_x$, which would not have to be very large if the positional noise $\Delta_u$ is small, since then $\xi_x$ is large)
will follow the horizontal locus labeled ``(1)" in Fig.~\ref{Fig5-TheoR}, and will pass directly from regime I to the unstable regime V. 
This is precisely the scenario we observe in our simulations of the apolar model.
Our numerical results show that for $L<L_{\rm c}$,  $S_1(L)$, $S_2(L)^{1/4}$, and $S_3(L)^{1/9}$ coincide (Fig \ref{Fig1-IsoRep}f), which is in full agreement with our theoretical prediction in region I, in which $S_n$ satisfy 
\begin{eqnarray}
S_n(L)\propto [S_1(L)]^{n^2}\, .
\label{snscalen2}
\end{eqnarray}
For $L\gg L_{\rm c}$,  $S_1(L)$, $S_2(L)^{1/4}$, and $S_3(L)^{1/9}$ depart from each other (Fig \ref{Fig1-IsoRep}f), and the undulation instability appears (Fig. \ref{Fig1-IsoRep}e), 
which implies the system is in the unstable region V. As mentioned earlier, the slow drift ``arches" of the undulations in our simulations
generates dislocations (Fig. \ref{Fig1-IsoRep}e), which destroys the smectic order. 
Therefore we expect to see a much faster algebraic decay of $S_n$ with $L$ in region V than in region I. 
This is, indeed, what we see, as shown in Fig. \ref{Fig1-IsoRep}f. Note also that the weak algebraic decay of $S_n$ for $L\ll L_{\rm c}$
observed in our simulations is {\it not} that predicted in \cite{live-soap}. That prediction applied in 
region IV in Fig. \ref{Fig5-TheoR}, which is the region of longest length scales in the {\it stable} case ($D_{ux}>0$).

%%%%%%%%%%%%%%%%%%%%%%%%%%%%%%%%%%%%%%%%%%%%%%%%%%%%%%%%%%%%%%%%%%%%%
\subsection{Polar smectic P}
%%%%%%%%%%%%%%%%%%%%%%%%%%%%%%%%%%%%%%%%%%%%%%%%%%%%%%%%%%%%%%%%%%%%%

Polar smectics P {\it do} have a different symmetry than polar smectic A, because the mean motion parallel to the layers 
(i.e., along the $x$-direction) breaks the $x\rightarrow-x$ symmetry.  
Adding to the original apolar model \eref{EOM_u} and \eref{EOM_rho} all $x$-inversion symmetry-breaking terms that are ``relevant", 
the hydrodynamic equations read \footnote{Despite the absence of $x\rightarrow-x$ symmetry, there is no contribution to $\partial_{t} u$ proportional to $\partial_{x} u$, since such a term violates {\it rotational} invariance.}:
\begin{eqnarray}
	\!\!\partial_{t} u & =&\!B\partial_{z}^2 u + D_{u x}\partial_x^2 u +g\partial_x^3 u- K  \partial_x^4 u+ C\partial_z\delta\rho+f_u \label{P_EOM_u}\,,\\ 
	\!\!\partial_{t} \delta\rho & =& v_\rho\partial_x \delta\rho+D_{\rho x}\partial_{ x}^2\delta\rho + D_{\rho z}\partial_z^2\delta\rho+ g_2\partial_x(\partial_z u)^2
  \nonumber\\
  &&+g_3\partial_z(\partial_xu\partial_zu)+g_4\partial_z(\delta\rho\partial_xu)+g_5\partial_x\left[\left(\delta\rho\right)^2\right]\nonumber\\
  &&+g_6\partial_x(\delta\rho\partial_zu) +v_2\partial_x\left[\partial_z u-\frac{1}{2}(\partial_x u)^2\right] +f_\rho  \,.
\label{P_EOM_rho}
\end{eqnarray}
The terms in these equations that are identical to those in (\ref{EOM_u}) and (\ref{EOM_rho}) have the same physical origin as they did there. The new terms (e.g., the $g$ term), are present here, but not in the apolar case, because they violate $x\rightarrow -x$ symmetry, which is {\it not} present in the polar case, but {\it is} present in the apolar case. 

The noises $f_u$ and $f_\rho$  have the same statistics \eref{fucor} and \eref{frhocor} as before.
As in the apolar case, the trivial ordered solution undergoes a long-wavelength instability at zero noise when $D_{ux}<0$. 
The dispersion relation for $q_z=0$ modes now reads $\nu(q_x)= -i g q_x^3 -D_{ux}q_x^2  - K  q_x^4$. 
Hence the instability (which again is strongest for $\bf{q}$ along $\bf{\hat{x}}$) 
has the same spatial structure as before, but propagates dispersively due to the $g$ term, with phase velocity $v_p(q_x, q_z=0)=gq_x^2$.
We thus expect (assuming as usual  that  nonlinear saturation does not modify the mode structure) 
the most unstable mode to propagate at a speed $v_{\rm c}$ given by $v_{\rm c}=v_p(q_x=q_c, q_z=0)=g q_{\rm c}^2={g|D_{u  x}|\over 2K}$. 

For $D_{ux}>0$, modes of any wavevector $\bq$ are now stable, but each propagates along the layers with the phase velocity $v_p(q_x, q_z=0)=gq_x^2$.
We observe such propagation of {\it fluctuations} in our simulations (Supp. Mat. Movie 2).

Linearizing Eqs. \eref{P_EOM_u} and \eref{P_EOM_rho}, we 
can calculate the behavior of the smectic order parameters $S_n$ with $L$ exactly 
as we did in the apolar case. Details are given in { \ref{App:hydroP}}; the result is that
the active polar smectic P, in this linear approximation, 
ultimately has the same crossover structure as that illustrated for the apolar case in Fig. \ref{Fig5-TheoR}. 
The only difference is that $ \eta_{_{_I}}=0$, which implies that in region I, the $S_n$'s, 
rather than falling off algebraically with $L$, are essentially constant. 

The above discussion was based entirely on the linear approximation. As mentioned earlier, the 
nonlinearities in Eqs. \eref{P_EOM_u}, \eref{P_EOM_rho} are relevant (technically, ``marginal") in 
$d=2$. We therefore expect that there exists a nonlinear length scale $L_{N\!L}$, growing 
very rapidly (specifically, like $\exp (A/\Delta^\nu)$, with $A$ a non-universal constant and $\nu$ a universal exponent that we have not yet determined) with decreasing noise strength $\Delta$,  where by $\Delta$ we mean a suitably weighted average of the noise strengths $\Delta_{u, \rho x, \rho z}$,
beyond which our theory is not valid.
What happens beyond $L_{N\!L}$ remains an open question, which can only be answered by a full renormalization group analysis. Since such an analysis is quite formidable, we restrict ourselves to plausible speculation.
The relevant nonlinearities may make the system much softer (i.e., fluctuate much more strongly) at long wavelength, as 
in other similar dynamical problems (e.g., the two dimensional KPZ equation \cite{KPZ}).
Then this softness will almost certainly lead to the unbinding of dislocations at the longest length scales $L\gg L_{N\!L}$, even when 
$D_{u x}$ is positive. This would imply that {\it all} polar smectic-P's would be disordered at the longest length scales.
Therefore, we expect both a breakdown of scaling law \eref{snscalen2} and a much faster algebraic decay of $S_n(L)$ with $L$ for $L\gg L_{N\!L}$. 
This  adds  another region to Fig. \ref{Fig5-TheoR}:
\begin{description}
\item[Region VI]: for polar smectic P with $L\gg L_{N\!L}$, nonlinear effects become important, which may induce dislocations and destroy smectic order. 
We expect $S_n(L)\propto L^{-2}$. 
\end{description} 

This new region (VI) may not exist for all polar smectics P. In active  polar smectics {\it A} \cite{Chen-Toner-2013},   nonlinear effects of the same type (i.e., marginal in $d=2$) as those found here destroy smectic order at long length scales in some, but not all, regions of parameter space. Whether  this can happen for polar smectics P remains an open question.

These predictions and speculations are in good agreement with our simulations, as illustrated in Fig.~\ref{Fig1-IsoRep}g, in which we again plot $S_1(L)$, $S_2(L)^{1/4}$, and $S_3(L)^{1/9}$.
They lie on top of each other for small system size, as predicted above for region I and IV. 
For large $L$ they depart from each other and decay much faster: 
we interpret the system size at which this happens as $L_{N\!L}$. 
That is, these simulations are following the locus labeled ``(2)'' in Fig.~\ref{Fig5-TheoR}. 
Apparently, the active tension $D_{ux}$ is so large that
the locus does not enter into regions II and III. 
However, unlike in the apolar case, $D_{ux}$  appears to be be positive 
here, since we observe no undulation instability. {Although, we emphasize, this is not a universal property of polar smectic-P order: it is entirely possible that microscopic models different from ours may exhibit an undulation instability.}

%%%%%%%%%%%%%%%%%%%%%%%%%%%%%%%%%%%%%%%%%%%%%%%%%%%%%%%%%%%%%%%%%%%%%
\section{Summary and discussion}
%%%%%%%%%%%%%%%%%%%%%%%%%%%%%%%%%%%%%%%%%%%%%%%%%%%%%%%%%%%%%%%%%%%%%

We have shown that active particle models which have %with 
both repulsion and alignment generically exhibit 
smectic configurations for sufficiently large densities. This occurs in both {\it apolar} models, 
in which the particles are equally likely to move in either direction parallel to their body axes, and in {\it polar} models, in which there is a preferred sign of motion along the axis. 

These results were obtained with Vicsek-style models, but we have checked that our conclusions also hold for continuous-time Langevin models of the type studied in \cite{ABP}; ergo, they are {\it not} artifacts of the discrete-time updating.
Depending on the symmetry of the particles and alignment, and on the anisotropy of the repulsion, 
smectic order and various dynamical large-scale phenomena emerge, as summarized in Table~\ref{table}.

We have found that a new type of smectic arrangement is most common
in our models: a state we call ``Smectic P'', 
in which particle alignment (and motion) is primarily {\it parallel} to the layers (Fig.~\ref{Fig1-IsoRep}b,c). 
This smectic-P order is unique to active particles: states with  the particle axes along the smectic layers have never been seen  in equilibrium.

In both the apolar and the polar cases, we observe weak algebraic decay of smectic order parameter $S_n(L)$ with system size $L$ for $L$ smaller than some crossover length, and much faster 
algebraic decay for larger $L$.  In the apolar case we also observe an undulation instability in large systems, while in the polar case we do not. 

To understand these phenomena, we have modified the hydrodynamic theories of active smectics A introduced in \cite{live-soap, Chen-Toner-2013} 
to treat both apolar and polar smectics P. The theory of  apolar smectics P
proves to be the same as that for apolar smectics A;  indeed, we interpret the instability we see as precisely the instability predicted in the ``negative active tension" case in \cite{live-soap}. We have, however, extended this theory of the apolar case to treat the regime of system sizes $L$ smaller than the instability length $L_c$. We find that two additional important length 
 scales smaller than $L_c$ can also appear for small values of the ``active tension" $D_{ux}$, 
 which is the most important fundamentally non-equilibrium parameter in the hydrodynamic theory. 
 This theory leads to the prediction of the five distinct regimes of behavior in 
($L,D_{ux}$)  plane illustrated in Fig. \ref{Fig5-TheoR}. 

The behavior that we observe in our simulations of the apolar case, as just summarized above, is entirely consistent with this hydrodynamic theory for a system with sufficiently negative active tension $D_{ux}$, such that, with increasing system size $L$, our system moves along the locus labeled (1) in Fig.~\ref{Fig5-TheoR}. Note that this instability is not inherent to all apolar systems. Some models that have the same symmetries as ours, but that differ in detail, may exhibit stable apolar phases. The hydrodynamic theory predicts that such a stable apolar system will exhibit algebraically slow decay of smectic order out to arbitrarily large system size, at least for sufficiently small noise. 

The hydrodynamic theory of the {\it polar} phase predicts rather different behavior. In addition to the five regions of the ($L,D_{ux}$) plane, a sixth region appears, which includes all systems larger than yet another characteristic length $L_{N\!L}$ (Fig.~\ref{Fig5-TheoR}).
We believe dislocations will always appear in this region, and make smectic order short ranged. 
Our numerical results are consistent with this prediction for a system with large positive active tension $D_{ux}$.

Note that, just as not all apolar systems need be unstable, nor do all polar systems need to be stable. 
Some polar models, different from ours, will probably exhibit an undulation instability of the type we observe in the apolar case, although, in the polar case, this instability be propagative.

We did {\it not} observe  ``giant" number fluctuations,  which are a 
signature of orientationally ordered but translationally disordered phases \cite{tonertuPRL1995, sradititonerEPL2003,GNF, Narayanan-jstat-2006} (Fig \ref{Fig4-SmGNF}).
But these simulations, because of  their intrinsic difficulty, were performed at sizes never much larger than the characteristic lengths $L_{\rm c}$ or $L_{N\!L}$.
We cannot exclude, and indeed believe, that giant number fluctuations, as well as 
other exotica generically present in orientationally-ordered flocks, will exist in very large active systems showing local smectic order and long-ranged orientational order.

Despite the general success of the hydrodynamic theories developed here in explaining many of the phenomena we observe in our simulations, 
some of them remain mysterious. In particular, the spontaneous rotation of the smectic layers 
(Fig.~\ref{Fig2-AniRep}a,b, Supp. Mat. Movies 4 \& 5)
is a type of spontaneous chiral symmetry breaking that is beyond the scope of the theories presented here, which assume an achiral steady state. 
We hope to develop a hydrodynamic theory of such symmetry breaking in future work.

\section*{Acknowledgments}
We thank F.\,Ginelli and S. Ramaswamy for lively and illuminating discussions
which took place within the Advanced Study Group 
``Statistical physics of collective motion'',  generously supported by the Max Planck Institute for the Physics of Complex Systems in Dresden. J.T. also thanks  the Aspen Center for Physics, Aspen, Colorado; the Isaac Newton Institute, Cambridge, U.K., and the Kavli Institute for Theoretical Physics, Santa Barbara, California; for their hospitality while this work was underway.  He also thanks the  US NSF for support by
awards \# EF-1137815 and 1006171; and the Simons Foundation for support by award \#225579. 
L.C. acknowledges support by the National Science Foundation of China (under Grant No. 11474354)

\appendix
%%%%%%%%%%%%%%%%%%%%%%%%%%%%%%%%%%%%%%%%%%%%%%%%%%%%%%%%%%%%%%%%%%%%%
\section{\label{App:hydroA}Hydrodynamic theory predictions for the apolar active smectic P phase
}
%%%%%%%%%%%%%%%%%%%%%%%%%%%%%%%%%%%%%%%%%%%%%%%

\subsection{Eigenfrequencies and Instability threshold
%apolar case
	\label{eig_inst}}
We begin by Fourier transforming the equations of motion \eref{EOM_u}, \eref{EOM_rho} in the main text:
\begin{eqnarray}
	\left[-i\omega+\Gamma_u({\bf q})\right]u({\bf q}, \omega)-iCq_z \delta\rho({\bf q}, \omega)&=&f_u({\bf q}, \omega)\label{uEOMFT} \\
	iq_z\Gamma_{\rho u}({\bf q}) u({\bf q}, \omega)+\left[-i\omega+\Gamma_\rho({\bf q})\right]\delta\rho({\bf q}, \omega)&=&f_\rho({\bf q}, \omega)\label{rhoEOMFT}
\end{eqnarray}
where we have defined: 
\begin{eqnarray}
&&\Gamma_u({\bf q})\equiv D_{u  x} q^2_ x  +K q_ x^4 + B q^2_z \, ,
\label{gudef}\\
&&\Gamma_\rho({\bf q})\equiv D_{\rho  x} q^2_ x  + D_{\rho z} q^2_z \, ,
\label{grhodef}
\end{eqnarray}
and
\begin{eqnarray}
\Gamma_{\rho u}({\bf q})\equiv C_ x q^2_ x  +C_z q^2_z \, .
\label{grhoudef}
\end{eqnarray}
The two eigenfrequencies $\omega_\pm$ of these equations of motion are, as usual, those values of $\omega$ that allow them to have non-zero solutions for $\delta\rho$ and $u$ when the forces $f_u$ and $f_\rho$ on the right hand sides are set to zero; i.e., they are the solutions of the eigenvalue equation:
\begin{eqnarray}
\omega^2+i\omega(\Gamma_u+\Gamma_\rho)+Cq_z^2\Gamma_{\rho u}-\Gamma_u\Gamma_\rho=0\, ,
\label{ev}
\end{eqnarray}
which are:
\begin{eqnarray}
\omega_\pm&=&-{i\over 2}\left[\Gamma_u({\bf q})+\Gamma_\rho({\bf q})\pm\sqrt{\left(\Gamma_u({\bf q})-
\Gamma_\rho({\bf q})\right)^2+4Cq_z^2\Gamma_{\rho u}({\bf q})}\right]\,.\label{ev2}
\end{eqnarray}
Stability requires that the imaginary part of these eigenfrequencies to be negative. The real part is readily seen to be zero.
This will be the case if and only if two conditions are satisfied:
\begin{eqnarray}
\Gamma_u({\bf q})+\Gamma_\rho({\bf q})>\sqrt{\left[\Gamma_u({\bf q})-
\Gamma_\rho({\bf q})\right]^2+4Cq_z^2\Gamma_{\rho u}({\bf q})}
\label{cond1}
\end{eqnarray}
and
\begin{eqnarray}
\Gamma_u({\bf q})+\Gamma_\rho({\bf q})>0 \, .
\label{cond2}
\end{eqnarray}
Squaring the first of these conditions (\ref{cond1}) implies
\begin{eqnarray}
\left(\Gamma_u({\bf q})+\Gamma_\rho({\bf q})\right)^2>\left(\Gamma_u({\bf q})-
\Gamma_\rho({\bf q})\right)^2+4Cq_z^2\Gamma_{\rho u}({\bf q})
\label{cond1.0}
\end{eqnarray}
which can be reorganized to read:
\begin{eqnarray}
\Gamma_u({\bf q})\Gamma_\rho({\bf q})-Cq_z^2\Gamma_{\rho u}({\bf q})>0\, .
\label{cond1.1}
\end{eqnarray}
Using our definitions (\ref{gudef}),  (\ref{grhodef}), and (\ref{grhoudef}) for $\Gamma_u({\bf q})$,
$\Gamma_\rho({\bf q})$, and $\Gamma_{\rho u}({\bf q})$ in this expression gives, after gathering terms,
\begin{eqnarray}
A q^4_ z  +F q_x^2q_z^2+D_{\rho  x} q^2_ x\left(D_{u  x} q^2_ x  +K q_ x^4\right)>0 \,,
\label{cond1.2}
\end{eqnarray}
where we have defined $A\equiv BD_{\rho  z}-CC_z$ and $F\equiv BD_{\rho  x}-CC_x+D_{u x}D_{\rho z}$.
Note that $A$ and $F$ will both be $>0$ for sufficiently small $C$, provided that $D_{u x}$ is not too large and negative. In this case
the left hand side of (\ref{cond1.2}) is a monotonically increasing function of  $q_z^2$, and, hence,
has its minimum at $q_z=0$. Hence, it is at $q_z=0$ that the condition (\ref{cond1.2}) will
first be violated. Furthermore, this violation will first occur with decreasing $q_x$ when $q_x=\sqrt{-D_{u x}/K}=q_{\rm max}$, where $q_{\rm max}$ is the critical value of $q_x$ defined in the main text. (Recall that here we are discussing the case $D_{ux}<0$.)
Thus, our claim in the main text that the first instability with increasing system size occurs at $q_z=0\, , q_x=q_{\rm c}$  follows provided that our second condition  (\ref{cond2}) is not violated at a larger $q$. But noting that 
$\Gamma_u({\bf q})$ and
$\Gamma_\rho({\bf q})$
are both $>0$ for any $q_x>q_{\rm c}$ (since then $D_{u  x} q^2_ x  +K q_ x^4>0$, and we also know that $B$ and both  $D_{\rho x}$ and $D_{\rho z}$ are $>0$), it is clear that (\ref{cond2}) can {\it not} be violated before (\ref{cond1}). This completes our demonstration that the first instability with increasing system size occurs, in the apolar case, at $q_z=0\, , q_x=q_{\rm c}$.

\subsection{\label{App:UuApolar}Fourier space $u-u$ correlation functions
% apolar case
}
The linear algebraic equations (\ref{uEOMFT}) and (\ref{rhoEOMFT}) are easily solved for  $u({\bf q}, \omega)$ and $\delta\rho({\bf q}, \omega)$. The result for $u({\bf q}, \omega)$ is
\begin{eqnarray}
u({\bf q}, \omega)={\left[-i\omega+\Gamma_\rho({\bf q})\right]f_u({\bf q}, \omega)+iCq_zf_\rho({\bf q}, \omega)\over(\omega-\omega_+)(\omega-\omega_-)}\, ,
\label{usol}
\end{eqnarray}
where $\omega_\pm$ are the two eigenfrequencies found above.
From equation (\ref{ev}),  we can read off
\begin{eqnarray}
\omega_++\omega_-=-i(\Gamma_u+\Gamma_\rho)\, ,
\label{evsum}
\end{eqnarray}
and
\begin{eqnarray}
\omega_+\omega_-=Cq_z^2\Gamma_{\rho u}-\Gamma_u\Gamma_\rho\, ,
\label{evprod}
\end{eqnarray}
both of which will prove useful later. We also note that $\omega_{\pm}$  are purely imaginary, which of course implies that $\omega_\pm^*=-\omega_\pm$.

Using this last fact, and our expression (\ref{usol}) for $u({\bf q}, \omega)$, we immediately get for the autocorelations
\begin{eqnarray}
	C_{uu}({\bf q}, \omega)&\equiv& \langle |u({\bf q}, \omega)|^2 \rangle \nonumber\\
			       &=&{ \langle |f_u({\bf q}, \omega)|^2 \rangle \left[\omega^2+\Gamma_\rho^2({\bf q})\right]
\over(\omega-\omega_+)(\omega-\omega_-)(\omega+\omega_+)(\omega+\omega_-)}\nonumber\\
&&+{\langle |f_\rho({\bf q}, \omega)|^2 \rangle
C^2 q_z^2\over(\omega-\omega_+)(\omega-\omega_-)(\omega+\omega_+)(\omega+\omega_-)}\, ,\label{Cuqo}
\end{eqnarray}
which is obviously born to be integrated by complex contour techniques. Before doing so, however, we need the spatio-temporally Fourier transformed autocorrelations of the noises
$\langle |f_u({\bf q}, \omega)|^2 \rangle$ and $\langle |f_\rho({\bf q}, \omega)|^2 \rangle $. These are easily read off from the real space correlation functions \eref{fucor},\eref{frhocor} in the main text:
\begin{eqnarray}
 &&\langle |f_\rho({\bf q}, \omega)|^2 \rangle = (\Delta_{\rho x} q_{ x}^2+\Delta_{\rho z} q_z^2)\equiv\Delta_\rho({\bf q}) \,  \\
&&\langle |f_u({\bf q}, \omega)|^2 \rangle = \Delta_u \, .
\end{eqnarray}
Using these in (\ref{Cuqo}) gives
\begin{eqnarray}
C_{uu}({\bf q}, \omega) &=&{\Delta_u\left[\omega^2+\Gamma_\rho^2({\bf q})\right]+(\Delta_{\rho x}q_ x^2+\Delta_{\rho z}q_z^2)
C^2 q_z^2\over(\omega-\omega_+)(\omega-\omega_-)(\omega+\omega_+)(\omega+\omega_-)}\, .
\label{uqo}
\end{eqnarray}

Integrating (\ref{uqo}) over all frequencies $\omega$ by obvious complex contour techniques then gives, after a little algebra,
the equal time correlation function:
\begin{eqnarray}
C^{ET}_{uu}({\bf q})&\equiv& \langle |u({\bf q}, t)|^2 \rangle \nonumber\\
&=&\int{d\omega\over 2\pi} \langle |u({\bf q}, \omega)|^2 \rangle
\nonumber\\&=&{-i\over2(\omega_++\omega_-)}\left\{\Delta_u-{\left[\Delta_u
\Gamma_\rho^2+C^2q_z^2\Delta_\rho({\bf q})\right]\over\omega_+\omega_-}\right\}\, ,
\label{uqt}
\end{eqnarray}
which can be simplified using our earlier expressions (\ref{evsum}) and (\ref{evprod}) for the sums and products of the eigenvalues, yielding, after a slight rearrangement of terms,  our final result:
\begin{eqnarray}
C_{uu}^{ET}({\bf q})&\equiv& \langle |u({\bf q}, t)|^2 \rangle
\nonumber\\&=&
{\Delta_u
\over 2\Gamma'_u({\bf q})}
+{C q_z^2\left[C\Delta_\rho({\bf q})-\Delta_u\Gamma_{\rho u}\right]\over2\left[\Gamma_u({\bf q})+
\Gamma_\rho({\bf q})\right]\Gamma'_u({\bf q})\Gamma_\rho({\bf q})}\, ,\nonumber\\&\equiv& C_{uu}^{ET(1)}+C_{uu}^{ET(2)}
\label{uqtsup}
\end{eqnarray}
where we have defined
\begin{eqnarray}
\Gamma'_u({\bf q})&\equiv& \Gamma_u({\bf q})  -Cq^2_z {\Gamma_{\rho u}({\bf q})\over\Gamma_\rho({\bf q})} \,\,\,\,\,,
\label{g'def}\\
C_{uu}^{ET(1)}
&\equiv&
{\Delta_u
\over 2\Gamma'_u({\bf q})}\,\,\,\,\,,
\label{c1q}
\end{eqnarray}
and
\begin{eqnarray}
C_{uu}^{ET(2)}({\bf q})\equiv{C q_z^2\left(C\Delta_\rho({\bf q})-\Delta_u\Gamma_{\rho u}\right)\over2\left[\Gamma_u({\bf q})+
\Gamma_\rho({\bf q})\right]\Gamma'_u({\bf q})\Gamma_\rho({\bf q})}\,.\,\,
\label{c2q}
\end{eqnarray}

The first term $C_{uu}^{ET(1)}$ in  (\ref{uqtsup}) exhibits very different behavior for wavevectors with $q_x\gg q_c\equiv\sqrt{|D_{u  x}|/2K}$ and $q_x\ll q_c$.  Note that for $D_{u x}<0$,  $q_c$ is the wavevector of maximum instability described in the main text. For $D_{u x}>0$,  $q_c$ is simply a crossover wavevector between the two regimes that we'll now describe.
For $q_x\gg q_{\rm c}$,
we can, by the definition of $q_c$,  neglect the $D_{ux}$ term in $\Gamma_u$.
Doing so, we immediately see that the first (i.e., the $\Gamma'_u$) term in the correlation function (\ref{uqt}) has the same form, for  wavevectors ${\bf q}$ with
$q_x\gg q_z$, as the full, equilibrium \cite{1st}
$C_{uu}({\bf q})$. It is given by
\begin{eqnarray}
C^{ET(1)}_{uu}({\bf q})={\Delta_u\over 2B'(q_z^2+\lambda^2 q_x^4)},  \label{CuuET1}
\end{eqnarray}
where we have defined the ``smectic penetration depth" $\lambda\equiv \sqrt{K/B'}$ and
%\begin{eqnarray}
$B'\equiv B-CC_x/D_{\rho x}$. Note that this is much larger than $q^{-2}$ for $q_x\gg q_z$; hence,
this range of wavevectors  dominates the contribution of $C^{ET}_{uu}({\bf q})$ to the real space fluctuations  for system sizes $L\ll L_{\rm c}$. Note also that $C^{ET}_{uu}({\bf q})$ is, for small $D_{u x}$, extremely anisotropic: for $q_x\gg q_c$, it scales like $q^{-2}$ for
%$q_z\sim q_x$,
$q_z\gg\lambda q_x^2$,
and like $q^{-4}$ for
%$q_z<\sim \lambda q_x^2$.
$q_z\ll \lambda q_x^2$.
For $q_x\ll q_c$, it is $\sim{\Delta_u\over B q^2}$ for $q_z\gg \sqrt{D_{ux}/B}q_x$, while it is $\sim{\Delta_u\over D_{u x} q^2}$  for $q_z\ll  \sqrt{D_{ux}/B}q_x$. In short, it is always much larger, for a given $|\bf{q}|$,  for small $q_z$; i.e., for directions of wavevector $\bf{q}$ near the plane of the smectic layers. It also changes its behavior dramatically between the range $q_x\ll q_c$, in which it scales like $q^{-2}$ for all directions of $q$,  and $q_x\gg q_c$, where it scales like $q^{-2}$ for $q_z\gg \lambda q_x^2$, and like $q^{-4}$ for
$q_z\ll \lambda q_x^2$.

The second term in (\ref{uqtsup}) has neither this anisotropy, nor this sensitivity to whether $q_x\ll q_c$ or $q_x\gg q_c$. To see this, note that the ratio ${Cq_z^2\over\Gamma'_u(\bf{q})}$ is
bounded above by ${C\over B'}$, while the sum $\Gamma_u({\bf q})+
\Gamma_\rho({\bf q})$ scales like $q^2$ for {\it all} directions of $\bf{q}$, even when $D_{u x}$ is small.  (Note that this is true even though we are considering $q_x\gg q_c$, which is, by definition,  the regime of wavevectors in which the $Kq_x^4$ term dominates the $D_{u x}q_x^2$ term, because the $D_{\rho x}q_x^2$ term in $\Gamma_\rho$ still dominates the $Kq_x^4$ term, since $D_{\rho x}$, unlike $D_{u x}$, need not be small when the activity is small.)

Indeed, this second term scales like $q^{-2}$ for all directions of $q$. We'll see in a moment that, as a result, this term leads to essentially isotropic algebraic decay of smectic order parameter correlations at all length scales.

Note also that only this second term grows with increasing density noise $\Delta_{\rho (x,z)}$ (since the first term depends only on the positional noise $\Delta_u$).

The above results are completely general. We will in what follows frequently consider the limit of large  $L_{\rm c}$, small positional noise $\Delta_u$, and large density noises $\Delta_{\rho (x,z)}$.
All three of these criteria are satisfied in our simulations: empirically, $L_{\rm c}\sim 100a$; the positional noise $\Delta_u$ should be proportional to the mean squared velocity fluctuations perpendicular to the layers, which, since the motion is primarily parallel to the layers, should scale like our angular noise strength parameter
$\sigma^2\sim10^{-4}$. Finally, the density noises $\Delta_{\rho (x,z)}$ are large, because our algorithm
introduces an $\cO(1)$ noise in the step that randomly, with equal probability, makes the particle move either forward or backward
along the direction selected.
This will not, in the small or zero $\sigma$ limit, contribute to the displacement noise $\Delta_u$, since in that limit all of the motion is along the layers.
But random statistical fluctuations in the number of particles moving left or right within a layer will clearly lead to fluctuations in the density; that is, to appreciable $\Delta_{\rho (x,z)}$.
This combination of large density noise and small positional noise means that
the second term $C_{uu}^{ET(2)}$ can, at intermediate length scales, actually dominate the behavior of the real space fluctuations of $u$, even though, as we'll  see in the next section, the contributions of the first term $C_{uu}^{ET(1)}$ grow more rapidly (for $L\ll L_{\rm c}$) as a function of distance.

%%%%%%%%%%%%%%%%%%%%%%%%%%%%%%%%%%%%%%%%%%%%%%%%%%%%

\subsection{Real space $u-u$ correlations, apolar case\label{real_uu}}

In order to predict the scaling with system size of the smectic order parameters we determine in our simulations, we need to calculate
$\langle|u({\bf r}+{\bf r}',t)-u({\bf r}\,',t)|^2\rangle\equiv \langle|\Delta u({\bf r})|^2\rangle$. This in general gets two contributions: a ``bulk" term given by:
\begin{eqnarray}
\langle|\Delta u({\bf r})|^2\rangle_B=2\int{d^2 q\over (2\pi)^2}
\left[1-\cos({\bf q}\cdot{\bf r})\right]C^{ET}_{uu}({\bf q}) \,  
\label{delugen}
\end{eqnarray}
and a ``zero mode" term $\du0$
given by:
\begin{eqnarray}
\langle|\Delta u({\bf r})|^2\rangle_0&=&{2\over L^2}\sum_{q_x\ne 0}
\left[1-\cos(q_x x)\right]
C^{ET}_{uu}(q_x, q_z=0,t)\,,
\label{deluzero1}
\end{eqnarray}
where the sum is, as usual, over all values  $q_x={2\pi m\over L}$, $m={\rm integer}$,  allowed by our periodic boundary conditions, excluding $m=0$.

The bulk term represents the contribution from all Fourier modes with $q_z\ne 0$, while the "zero mode" term, as its name suggests, incorporates the contribution from modes with $q_z=0$.
The latter modes would be absent in a system with fixed boundary conditions (i.e., $u=0$ on the boundaries), but is present in our simulations, since we use periodic boundary conditions.

We'll consider first the contribution of the first term $C_{uu}^{ET(1)}$ to $\dut$, which we'll call $\du1$, and then that of the second term $C_{uu}^{ET(2)}$, which we'll equally unimaginatively call
$\du2$.

Let's first consider the case $L\ll L_c$. In this case we are in the regime $x\ll L_{\rm c}$ and $z\ll  {L_{\rm c}^2\over\lambda}$.
In this regime, the range of wavevectors $q_x\gg q_c$, $q_z\ll q_x$ dominates the contribution of the first
term to both the bulk term (\ref{delugen}) and the zero mode piece (\ref{deluzero1}).
We can therefore, for this range of $\bf{r}$, replace the first term in (\ref{uqtsup}) with Eq. (\ref{CuuET1}),
which, as noted earlier, has {\it exactly} the same form as in an equilibrium smectic, for which these correlations were computed long ago \cite{1st}. Hence,
for the contribution
from the bulk term from $\cqf$, we can simply quote the results of \cite{1st} with the trivial replacement of $k_BT$ in the equilibrium problem with $\Delta_u/2$; this gives 
\begin{eqnarray}
\langle|\Delta u({\bf r})|^2\rangle_{1,B}&=&{2|x|\over q_0^2 \xi^B_x}g\left({x^2\over \lambda |z| }\right)
=\left\{
\begin{array}{ll}
{2|x|\over q_0^2\xi^B_x}\,\,\,\,\,\,\,\,\,,
&|x|\gg \sqrt{\lambda |z|}\,,
\\ \\
{2\over q_0^2}\sqrt{{ |z|\over\xi_z}}\,,
&|x|\ll \sqrt{\lambda |z|}\,,
\end{array}\right.
\label{deluapolsup}
\end{eqnarray}
where we have defined the correlation lengths
\begin{eqnarray}
\xi^B_x\equiv{8\sqrt{B'K}\over q_0^2\Delta_u}\,,
\label{xi_x}
\,\,\,\,\,\,\,\,\,\,\,\,\,\,\,\,{\rm and}
\,\,\,\,\,\,\,\,\,\,\,\,\,\,\,\,
\xi_z\equiv{\pi(\xi^B_x)^2\over 4\lambda}\,,
\label{xi_z}
\end{eqnarray}
along $x$ and $z$ respectively, and the scaling  function $g(w)$  as a function of its argument $w\equiv{x^2\over \lambda |z| }$ is \cite{1st}
\begin{eqnarray}
g(w)={\rm erf}(\sqrt{w}/2)+{e^{-w/4}\over\sqrt{ 4\pi w}}\,.
\label{eqscale}
\end{eqnarray}

The lengths $\xi^B_x$ and $\xi_z$ are the distances along $x$ and $z$ at which the rms fluctuations $\sqrt{\du{1,B}}$ are of order $a_l$ (more precisely, they are ${\sqrt{2}\over 2\pi}a_l$).
Note that  $\xi_z\gg\xi^B_x$ in well ordered systems where $\xi^B_x\gg\lambda$.

In most systems, the contribution of the zero modes are
negligible, due to the factor of $1\over L^2$ in front of Eq. (\ref{deluzero1}). In active smectics, however, if the system size $L \ll L_{\rm c}$, the critical size above which active tension effects become important, so that we can drop the active tension term $D_{u x}q_x^2$ relative to the $Kq_x^4$ term (since, for such system sizes, $q_x\ll q_c\equiv\sqrt{D_{u x}/K}$), $C_{uu}(q_x, q_z=0,t)$ is so large at small $q_x$ (diverging like ${1\over q_x^4}$ as $q_x\rightarrow 0$), that this ``zero mode" term can actually dominate.  Indeed, for $L\ll L_{\rm c}$, we find that, for $x\ll L$, the sum in (\ref{deluzero1}) is dominated by the smallest allowed $q_x$'s, for which we can expand the cosine to leading order in ${x\over L}$. This gives 
\begin{eqnarray}
\langle|\Delta u({\bf r})|^2\rangle_{1,0}&\approx&{1\over L^2}\sum_{q_x\ne 0}
(q_x x)^2C_{uu}(q_x, q_z=0,t)
\nonumber\\
&=&2\left({x\over L}\right)^2\sum_{m=1}^\infty\left({2\pi
m\over L}\right)^2{\Delta_u\over  K\left({2\pi m\over L}\right)^4}\nonumber\\
&\equiv&{2\over q_0^2}\left({x\over  \xi_x}\right)^2\,,
\label{deluzero2}
\end{eqnarray}
where we have defined a correlation length  $\xi_x$ given by
\begin{eqnarray}
\xi_x={4\over q_0}\sqrt{{3K\over \Delta_u}}\,
 \label{xi0}
 \end{eqnarray}
 such that when  $x=\xi_x$ the zero mode real space fluctuations $\dut_{_0}={2\over q_0^2}$, where $q_0\equiv{2\pi\over a_l}$, with $a_l$ the layer spacing.
This correlation length is clearly much shorter than that coming from the ``bulk modes", which is given by Eq. (\ref{xi_x}), in the small noise ($\Delta_u\rightarrow 0$) limit, since it diverges like $1/\sqrt{\Delta_u}$ as $\Delta_u\rightarrow 0$, while the bulk correlation length
 diverges like $1/\Delta_u$ as $\Delta_u\rightarrow 0$. Hence, we expect $\xi_x$ to give the correlation length in the $x$ direction for small displacement noise $\Delta_u$.

This zero mode contribution (\ref{deluzero2}) dominates the bulk contribution (\ref{deluapolsup}) for $|x|\ll\sqrt{\lambda |z|}$, provided that ${x^2\Delta_u\over24K}\gg{2\over q_0^2}\sqrt{{ |z|\over\xi_z}}$. This leads to the requirement $|x|\gg\sqrt{{48K\over q_0^2\Delta_u}}\left({|z|\over\xi_z}\right)^{1/4}$, which is clearly much smaller, for $|z|\gg 36 \lambda$, than the value of $x$ at which the bulk term crosses over from being controlled by $z$ to being controlled by $x$; i.e., $|x|\ll\sim{\lambda |z|}$. Hence, for all large $r$'s at which the bulk term makes an appreciable contribution to $\du1$, we can replace that bulk term by
${2\over q_0^2}\sqrt{{ |z|\over\xi_z}}$. Hence, we can {\it always} (for $L\ll L_{\rm c}$) and large $r$ replace
$\du1=\du{1,B}+\du{1,0}$ with
\begin{equation}
\langle|\Delta u({\bf r})|^2\rangle_1 =
\frac{2}{q_0^2}\left( \sqrt{\frac{|z|}{\xi_z}}+\left(x\over\xi_x\right)^2\right)\,.
\label{deluapol}
\end{equation}

Now we consider the case  $L\gg L_c$. In this case the zero modes' contribution to $\langle |\Delta u({\bf r})|^2\rangle$ is always negligible. For $x\ll L_c$ and $z\ll {L_c^2\over\lambda^2}$, $\langle|\Delta u({\bf r})|^2\rangle_{1,B}$ is again given by Eq. (\ref{deluapolsup}). For $x\gg L_c$ or $z\gg {L_c^2\over\lambda^2}$, the integration of $C_{uu}^{ET(1)}(\bf q)$ over large $\bf q$'s  (i.e., $q_x\gg q_c$) gives a constant, which can be obtained by evaluating Eq. (\ref{deluapolsup})
at $x=L_c, z=0$.

It is the integration  of $C_{uu}^{ET(1)}$ over small $\bf q$'s (i.e., $q_x\ll q_c$) that dominates in this regime of $L$.
In this regime of $\bf q$',s in the stable case $D_{u x}>0$,
$\Gamma_u\,'$ scales like $q^2$ for all directions of ${\bf q}$. Specifically,
in this limit, $\Gamma_u\,'\approx B'q_z^2+D_{u x}q_x^2$. Using this in (\ref{c1q}), and using the result in (\ref{delugen}) plus the constant
then gives
\begin{eqnarray}
\langle|\Delta u({\bf r})|^2\rangle_1={\Delta_u\over4\pi\sqrt{BD_{u x}}}\ln\left(R\over L_c\right)+{2L_c\over q_0^2 \xi^B_x}
\label{deluapol3}
\end{eqnarray}
where $R\equiv\sqrt{x^2+\left(D_{u x}\over B\right)z^2}$.

Now let's consider the contribution of the second term in
(\ref{uqtsup}) to the real space fluctuations of $u$. Since this scales like $1/q^2$ for all directions of $\bf{q}$, it will also make contributions to the mean squared  real space fluctuations $\langle|\Delta u({\bf r})|^2\rangle$ that scale like $\langle|\Delta u({\bf r})|^2\rangle_2\propto\ln{\left({r\over a_l}\right)}$. This scaling holds for this term for all $r$. The  detailed calculation goes as follows:
we start by rewriting $\cqs$ in polar coordinates $(q, \theta)$ where $(q_x, q_z)=q(\sin\theta, \cos\theta)$:
\begin{eqnarray}
{\cqs}=
{C\over 2q^2}\tau(\theta)\,\,\,,
\label{c2pol}
\end{eqnarray}
where we have defined: 
\begin{eqnarray}
\tau(\theta)\equiv{\left(F\sin^2\theta+G\cos^2\theta\right)\over \left(\bar{D}_x\sin^2\theta+\bar{D}_z\cos^2\theta\right)\left(J\sin^2\theta+H\cos^2\theta\right)}
\,\,\,,
\label{taudef}
\end{eqnarray}
\begin{eqnarray}
F\equiv
C\Delta_{\rho x}-C_x\Delta_u\,\,\,,
\label{Fdef}
\end{eqnarray}
\begin{eqnarray}
G\equiv
C\Delta_{\rho z}-C_z\Delta_u\,\,\,,
\label{Gdef}
\end{eqnarray}
\begin{eqnarray}
J\equiv B'D_{\rho x}\,\,\,,
\label{Jdef}
\end{eqnarray}
\begin{eqnarray}
H\equiv B''D_{\rho z}\,\,\,.
\label{Hdef}
\end{eqnarray}
\begin{eqnarray}
B''\equiv B-CC_z/D_{\rho z}\,\,\,,
\label{B''}
\end{eqnarray}
\begin{eqnarray}
\bar{D}_x\equiv D_{\rho x}+D_{u x}\,\,,
\label{Dbarx}
\end{eqnarray}
and
\begin{eqnarray}
\bar{D}_z\equiv D_{\rho z}+B\,\,.
\label{Dbarz}
\end{eqnarray}
Inserting (\ref{c2pol}) into the general expression (\ref{delugen}) for $\dut$ gives 
\begin{eqnarray}
	\langle|\Delta u({\bf r})|^2\rangle_2&=&2\int{d^2 q\over (2\pi)^2}
\left[1-\cos({\bf q}\cdot{\bf r})\right]C^{ET(2)}_{uu}({\bf q})\nonumber\\&=&{C\over 2\pi^2}\int_{-{\pi\over 2}}^{{\pi\over 2}}  \tau(\theta)d\theta\int_0^\Lambda dq\,{\left[1-\cos({\bf q}\cdot{\bf r})\right]\over q}\,\,\,,
\label{delu2.1}
\end{eqnarray}
where $\Lambda$ is an ultraviolet cutoff of order ${1\over a_l}$.

The integral over $q$ in this expression
can readily be evaluated for large $r$ by dividing it, like Gaul\footnote{Gaul refers to a historic geographic region in Europe, roughly contiguous with present-day France.} , into three parts:
\begin{eqnarray}
\int_0^\Lambda dq\,{\left[1-\cos({\bf q}\cdot{\bf r})\right]\over q}=I-II+III\,\,\,,
\label{int.1}
\end{eqnarray}
where we have defined
\begin{eqnarray}
I\equiv\int_{1\over r|\cos(\theta-\phi)|}^\Lambda{ dq\over q}=\ln(\Lambda r)+\ln(|\cos(\theta-\phi)|)\,, 
\label{I.1}
\end{eqnarray}
\begin{eqnarray}
II\equiv\int_{1\over r|\cos(\theta-\phi)|}^\Lambda dq\,{\cos( qr|\cos(\theta-\phi)|)\over q}\,\,\,,
\label{II.1}
\end{eqnarray}
and
\begin{eqnarray}
III\equiv\int_0^{1\over r|\cos(\theta-\phi)|} dq\,{\left[1-\cos(qr|\cos(\theta-\phi)|)\right]\over q}\,.  
\label{III.1}
\end{eqnarray}
In these expressions, $\phi$ is the angle
between ${\bf r}$ and the $z$-axis, and we have exploited the evenness of
the cosine to replace its argument with its absolute value.

The integral $II$ clearly converges as $q\rightarrow\infty$, due to the oscillation of the cosine, combined with the ${1\over q}$ falloff. We can therefore safely take the upper limit on this integral to $\infty$ for large $r$, where the lower limit is small, and obtain
\begin{eqnarray}
II\approx\int_{1\over r|\cos(\theta-\phi)|}^\infty dq\,{\cos( qr|\cos(\theta-\phi)|)\over q}\,\,\,.
\label{II.2}
\end{eqnarray}
Now, making the linear change of variables of integration from $q$ to   $\ell\equiv qr|\cos(\theta-\phi)|$ in both (\ref{II.2}) and (\ref{III.1}) gives
\begin{eqnarray}
II\approx\int_1^\infty d\ell\,{\cos \ell\over \ell}\,\,\,,
\label{II.3}
\end{eqnarray}
and
\begin{eqnarray}
III=\int_0^1 d\ell\,{[1-\cos \ell]\over \ell}\,\,\,,
\label{III.3}
\end{eqnarray}
both of which are clearly finite, $\cO(1)$ constants, independent of $r$ and $\phi$. Using this fact, and (\ref{I.1}), in (\ref{int.1}), we obtain
\begin{eqnarray}
\int_0^\Lambda dq\,{\left[1-\cos({\bf q}\cdot{\bf r})\right]\over q}&=&\ln(\Lambda r)+\ln(|\cos(\theta-\phi)|)+\cO(1)\nonumber\\&=&\ln({r\over a_l})+\ln(|\cos(\theta-\phi)|)+\cO(1),
\label{int.2}
\end{eqnarray}
where in the second equality we have absorbed a constant $\ln(\Lambda a)$ into the other $\cO(1)$ constants coming from the integrals $II$ and $III$.

Using this result in (\ref{delu2.1}) gives
\begin{eqnarray}
\langle|\Delta u({\bf r})|^2\rangle_2 &=&{C\over 2\pi^2}\int_{-{\pi\over 2}}^{{\pi\over 2}}  \tau(\theta)d\theta\left(\ln({r\over a_l})+\ln(|\cos(\theta-\phi)|)+\cO(1)\right)\nonumber\\
&=&{C\Upsilon\over4\pi}\left[\ln{\left({r\over a_l}\right)}+h(\phi)+\cO(1)\right]
\,\,\,,
\label{delu2.2}
\end{eqnarray}
where we have defined
\begin{equation}
\Upsilon\equiv
%\nonumber\\
{2\over\pi}\int_{-{\pi\over 2}}^{{\pi\over 2}}  \tau(\theta)d\theta\,\,\,,
\label{Upsdef}
\end{equation}
and
\begin{eqnarray}
h(\phi)\equiv
{2\over\pi\Upsilon}\int_{-{\pi\over 2}}^{{\pi\over 2}}  \tau(\theta)\ln(|\cos(\theta-\phi)|)d\theta\,\,\,,
\label{hdef}
\end{eqnarray}

The integral for $\Upsilon$ can be evaluated by the simple trigonometric substitution $v\equiv \tan\theta$, which gives
\begin{eqnarray}
\Upsilon\equiv
%\nonumber\\
{2\over\pi}\int_{-\infty}^{\infty} dv  {\left(Fv^2+G\right)\over \left(\bar{D}_xv^2+\bar{D}_z\right)\left(Jv^2+H\right)}\,\,\,,
\label{Ups1}
\end{eqnarray}
which can be straightforwardly evaluated by complex contour techniques, giving:
\begin{eqnarray}
\Upsilon&=&
	{2\left(C\Delta_{\rho z}-C_z\Delta_u\right)\over B''D_{\rho z}\sqrt{\bar{D}_x\bar{D}_z}+\bar{D}_z\sqrt{B'B''D_{\rho x}D_{\rho z}}}\nonumber\\&&+{2\left(C\Delta_{\rho x}-C_x\Delta_u\right)\over B'D_{\rho x}\sqrt{\bar{D}_x\bar{D}_z}+\bar{D}_x\sqrt{B'B''D_{\rho x}D_{\rho z}}}\,\,\,.
\label{Ups}
\end{eqnarray}
%with
%
One might think that this logarithmic divergence  would, for $r\ll L_c$, be dominated by the stronger power law divergence of the contributions $\langle|\Delta u({\bf r})|^2\rangle_1$ coming from the first term, at least for large $r$. But  if the density noises $\Delta_{\rho (x,z)}$ are large and the positional noise $\Delta_u$ is small, this need not be the case,
since the coefficient $\Upsilon$ of the logarithmic divergence has pieces that grow linearly with $\Delta_{\rho x,z}$. And we do in fact expect that, in our simulations, the density noises $\Delta_\rho$ are  large while the
displacement noise $\Delta_u$ is small, as explained in the main text.

So, in the limit of small angular noise, we expect $\Delta_{\rho (x,z)}$ to be large, by some suitable dimensionless measure, compared to $\Delta_u$.
This means that the contribution from the non-equilibrium second term in (\ref{uqt}) to $\dut$ can actually dominate that of the first term all the way out to the correlation length (i.e., $\xi_x$ given by (\ref{xi0}) for $L\ll L_c$, or $\xi_x^B$ given by (\ref{xi_x}) for $L\gg L_c$) provided that the value of $\du2$ is greater than ${2\over q_0^2}$ (the value of the contribution of the first term at the correlation length)
there. This leads to the condition
\begin{eqnarray}
{C\Upsilon\over4\pi}\ln{\left(X\over a_l\right)}\gg{2\over q_0^2}\,,
 \label{delrhobound1}
\end{eqnarray}
where $X$ stands for either $\xi_x$ or $\xi_x^B$.
Assuming, as seems reasonable, that the  diffusion constants $\bar{D}_{x,z}$ and
$D_{\rho (x,z)}$ are all comparable, and are larger than or comparable to $D_{u x}$, and likewise assuming that $B''\sim B'$ and $\Delta_{\rho x}\sim\Delta_{\rho z}\equiv \Delta_\rho$, we can estimate $\Upsilon\sim {C \Delta_\rho\over B'D^2}$, where
$D$ is the common
value of the diffusion constants $\bar{D}_{x,z}$ and
$D_{\rho (x,z)}$. On dimensional grounds, we expect  $D\sim\rho_0 C$, where $\rho_0$ is the mean density. Using this estimate, we obtain  a lower bound  on the density noise required for the second, non-equilibrium term in (\ref{uqt}) to dominate all the way out to the correlation lengths $\xi_x$ and $\xi_x^B$:
\begin{eqnarray}
\Delta_\rho\gg{8\pi B'\rho_0^2\over  q_0^2\ln{\left(X\over a_l\right)}}\,,
 \label{delrhobound2}
\end{eqnarray}
which should be satisfied in our simulations since $\Delta_\rho$ is large, for the reasons discussed earlier, while $\Delta_u$ is small, which makes {both $\xi_x$ and $\xi_x^B$ large, thereby reducing the right hand side of (\ref{delrhobound2}).

In summary, for values of $\Delta_\rho$ that are large enough, in the sense just described, and
values of the active tension $D_{u x}$ that are small enough, there are four important crossover
lengths. The first is $L_{\rm c}$, beyond which the effects of the active tension become
important. The other three, which are independent of $D_{u x}$ in this limit, are the correlation lengths $\xi_x$, $\xi_x^B$ (for $L\ll L_c$ and $L\gg L_c$, respectively),
and $\xi_z$
that give the distances along $x$ and $z$ beyond which $\dut$ grows algebraically with
$x$ and $z$. These length scales divide the $D_{u x}$-system-size $L$ plane into five regions, as
illustrated  in Fig. \ref{Fig5-TheoR}. The active tension $D_{u x}$ is only important for length
scales $L\gg L_{\rm c}\equiv\sqrt{K/|D_{u x}|}$, which defines the upper and lower bounding curves.
Because of this, regions I, II, and III extend right across the $L$-axis, where $D_{u x}=0$. In region
I ($L\ll\xi_x$), the second, entirely non-equilibrium, logarithmic  contribution $\du2$ to $\dut$
dominates. In region II ($\xi_x\ll L\ll\xi_z$), $\sqrt{\dut}$ is much bigger than the lattice spacing $a_l$
for $\bf{r}$ primarily along $\hat{\bf{x}}$ (i.e., the layer direction) and $x\gg \xi_x$, but is much less
than $a_l$ for all points whose separation lies primarily in the $\hat{\bf{z}}$ direction (i.e., the layer
normal). Region IV ($D_{u x}>0$ and $L\gg L_{\rm c}$) is not further divided into more sub-regions by the correlation lengths $\xi_x$ and $\xi_z$, since the behavior of $\dut$ is always logarithmic for large $r$ (i.e., $x\gg L_c$ or $z\gg L_c^2/\lambda$) in all directions, which solely determines the $L$-dependence of the quantitiy $S_n$. And finally, for $D_{u x}<0$ and $L\gg L_{\rm c}$, the smectic state is unstable. We do not have any analytic theory for the long-term, large distance state of the system in this regime. However, as argued in the main text, we expect the instability to induce nucleation of dislocations, which will cut off order prameter correlations at $L_{\rm c}$, hence, for $L\gg L_{\rm c}$, we expect $S_n\propto{1\over A}={1\over L^2}$.

To summarize the above rather complicated discussion,
$\dut=\du1+\du2$,  where $\du{1,2}$ represent respectively the contributions of the first and second terms in \eref{uqt}.  For $L\ll L_c$, $\du1$ is given by
\begin{equation}
\langle|\Delta u({\bf r})|^2\rangle_1 =
\frac{2}{q_0^2}\left( \sqrt{\frac{|z|}{\xi_z}}+\left(x\over{\xi_x}\right)^2\right)\,;
\label{deluapolL<Lc}
\end{equation}
while for $L\gg L_c$, it is  given by
\begin{equation}
\langle|\Delta u({\bf r})|^2\rangle_1 =
\cases{
		\frac{2}{q_0^2}\left(\sqrt{\frac{|z|}{\xi_z}}+\frac{|x|}{\xi_x^B}\right) & $|x|\ll L_c, |z| \ll L_c^2/\lambda$  \\
        	\frac{\Delta_u\ln{\left(\frac{R}{L_c}\right)}}{4\pi\sqrt{BD_{u x}}}+\frac{2 L_c}{q_0^2\xi_x^B} &   $|x|\gg L_c \textnormal{ or } |z| \gg L_c^2/\lambda$
	}
\label{deluapolL>Lc}
\end{equation}
where $R\equiv\sqrt{x^2+\left(D_{u x}\over B\right)z^2}$.
And $\du2$ is always given by
\begin{eqnarray}
\langle|\Delta u({\bf r})|^2\rangle_2 = {C\Upsilon\over4\pi}\ln{\left({r\over a_l}\right)}+h(\phi)+\cO(1)\,,
\label{delu2s}
\end{eqnarray}
regardless of whether $L\ll L_c$ or $L\gg L_c$. We remind the reader that $\phi$ is the angle between ${\bf r}$ and the $z$-axis$, \Upsilon$ is a linear function of the noises $\Delta_{\rho (x,z)}$ and $\Delta_u$ given by Eq. (\ref{Ups}), and the correlation lengths are given by $ \xi_x={4\over q_0}\sqrt{{3K\over \Delta_u}}$,
$\xi_z={16\pi B'^2\lambda\over q_0^4 \Delta_u^2}$, and $\xi^B_x={8\sqrt{B'K}\over q_0^2\Delta_u}$ .
Note that $\xi_z\gg \xi_x$ for small $\Delta_u$.

\subsection{Smectic Order parameters, apolar case\label{smop_apolar}}

We'll now discuss the implications of these results for the smectic order parameters that we measure in our simulations.

As in equilibrium, smectic order is characterized by  an infinite set of  local complex smectic order parameters  $\psi_n({\bf r},t)$ defined via \cite{deGennes}
\begin{eqnarray}
\rho({\bf r},t)\equiv\rho_0+\sum_{n=1}^\infty\psi_n({\bf r}\,,t)e^{inq_0z}+\rm{c.c.}\,,
\label{psidefsup}
\end{eqnarray}
where $\rho$ is the number density, $\rho_0$ its mean, and $q_0\equiv2\pi/a_l$, with $a_l$ the distance between neighboring layers.

To quantify the order in our simulations, we numerically determine the quantities
$S_n\equiv I({\bf q}_n,t)$, where the ``intensity"\footnote[2]{So-called, because it is proportional to the
intensity of scattering of electromagnetic radiation by the smectic when the wavelength of that radiation is comparable to the smectic layer spacing.}
 $I({\bf q},t)\equiv{|\rho({\bf
q},t)|^2\over N^2}$, with $N=\rho_0 A$ the number of particles in the system, $A=L^2$ is the area of the system, ${\bf q}_n\equiv nq_0
\hat{z}$, and
\begin{eqnarray}
 \rho ({\bf q},t)=\int d^2r \, \rho({\bf r},t)e^{-i{\bf q}\cdot{\bf r}}\,.
\end{eqnarray}

In writing the expansion (\ref{psidefsup}), we implicitly assume that the $\pnrt$ are slowly varying in space and time (i.e., that they only have support in Fourier space at small wavevectors and frequency). That is, we have absorbed the rapid variation of the density in the smectic phase into the complex exponentials $e^{inq_0z}$, while the $\pnrt$ embody the slow spatial variations of the positions $\urt$ of the layers. Given this,
the  definition (\ref{psidefsup}) is equivalent to:
\begin{eqnarray}
\rnqo=\pno\,\,,
\label{psidef2}
\end{eqnarray}
from which it follows from our definition of the $S_n$'s that
\begin{eqnarray}
S_n={\langle |\pno|^2 \rangle\over N^2}
&=&{\int d^2rd^2r' \langle \psi_n^*({\bf r}, t)\psi_n({\bf r}\,', t) \rangle \over N^2}\nonumber \\
&=&{\int d^2 r  \langle \psi_n^*({\bf r}, t)\psi_n({\bf 0}, t) \rangle \over \rho_0^2 A} \,\,\,.
\label{OPsup}
\end{eqnarray}

The phases of these local order parameters are all proportional to the displacement field $u$; that is, we can write  $\pnrt=|\pnrt|e^{-inq_0u({\bf r}\,,t)}$. Hence, since the amplitudes $|\pnrt|$ of the smectic order parameters are {\it not} Goldstone modes of the system, and are therefore not expected to have large fluctuations, the decay of the of the correlation function $\langle \psi_n^*({\bf  r}, t)\psi_n({\bf 0}, t) \rangle $ with ${\bf  r}$ is driven primarily by the fluctuations of the layer displacement $\urt$. Motivated by this, we replace $|\pnrt|$ with a real positive constant $\psi_n^0$, and obtain for the correlation function we need:
\begin{eqnarray}
\langle \psi_n^*({\bf r}, t)\psi_n({\bf 0}, t) \rangle =(\psi_n^0)^2\langle \exp\left[-inq_0\Delta u({\bf r})\right] \rangle \,\,\,.\label{psicor1}
\end{eqnarray}
where $\Delta u({\bf r})\equiv u({\bf r},t)-u({\bf 0},t)$.

This can be related to the mean squared real space fluctuations  $\langle|\Delta u({\bf r})|
^2\rangle$
%given in equation (\ref{deluapol})
by noting that $\Delta u({\bf r})$ is a zero-mean Gaussian random variable. This follows from the fact that it is a linear function of the
Fourier components $u({\bf q}, \omega)$,  which are  in turn linear functions of the Fourier
components $f_u({\bf q}, \omega)$ and  $f_\rho({\bf q}, t)$. Since these noises are themselves,
by assumption, zero-mean Gaussian random variables, so is $\Delta u({\bf r}-{\bf r}\,')$.

Using this fact, we can use the well-known (and easily derived by any who don't know it) relation for any
zero-mean Gaussian random variable $x$ that $\langle e^{ix}\rangle=\exp(-{1\over 2}\langle x^2\rangle)$ to obtain
\begin{eqnarray}
\langle \exp\left[-inq_0\Delta u({\bf \Delta r})\right] \rangle =\exp\left(-{n^2q_0^2\over 2}\langle|\Delta u({\bf \Delta r})|^2\rangle\right)\,.\label{psicor2}
\end{eqnarray}
Using this in (\ref{psicor1}), and using (\ref{psicor1}) in (\ref{OPsup}) gives
\begin{eqnarray}
S_n=w_n{\int_0^Ldx\int_0^Ldz \exp\left(-{n^2q_0^2\over 2}\langle|\Delta u({\bf  r})|^2\rangle\,,\right)\over A}\,, \label{snscale1sup}
\end{eqnarray}
as claimed in the main text \eref{snscale1},
where we have defined
\begin{eqnarray}
w_n\equiv\left({\psi_n^0\over\rho_0}\right)^2.\label{Wn}
\end{eqnarray}

Using  our earlier results (\ref{deluapolL<Lc}),  (\ref{deluapolL>Lc}), and (\ref{delu2s}) for
$\dut$ in  (\ref{snscale1sup}), we see that, for $L\ll L_{\rm c}$,
\begin{eqnarray}
\!S_n &=&\frac{w_n}{A} \!\int_0^L\!\!\!dx \!\int_0^L\!\!\!dz \left({r\over a_l}\right)^{-\eta_In^2} \exp\left[-n^2\!\left( \sqrt{\frac{|z|}{\xi_z}}
+\frac{x^2}{\xi_x^2}\right.\right.\left.\left.+{q_0^2h(\phi)\over 2}\right)\!\right]
\label{snscale2}
\end{eqnarray}
with
\begin{equation}
\eta_I\equiv{Cq_0^2\Upsilon\over8\pi}
\label{eta1}
\end{equation}
For $L\ll \xi_x$ and $L\ll L_{\rm c}$, denoted as region I in Fig. \ref{Fig5-TheoR}, the exponential factor in this
expression is nearly $1$ over the entire region of integration (recall that $\xi_z\gg\xi_x$), and, as a result,
the $S_n$'s fall off algebraically with $L$: $S_n\propto L^{-n^2\eta_I}$.
For  $\xi_z\gg L\gg \xi_x$ and  $L\ll L_{\rm c}$, denoted as region II in Fig. \ref{Fig5-TheoR},
we can ignore the $z$-dependence of the exponential factor, but not the $x$-dependence.
The integral is then clearly dominated by $z\sim L\gg \xi_x$ and $x\sim\xi_x$, which means $x\ll z$ in this dominant region.
We can therefore replace $r$ with $z$ and $\phi$ with $0$ in \eref{snscale2}, and obtain
\begin{eqnarray}
S_n &\approx&  \frac{w_n}{A} \!\! \int_0^L\!\!\!dx \!\!\int_0^L\!\!\!dz \exp\left[-\frac{n^2x^2}{\xi_x^2}-{n^2q_0^2h(0)\over 2}\right]\!\left(\frac{z}{a_l}\right)^{-\eta_I n^2} \!\!\! \nonumber \\ 
&\propto&  \frac{L^{-1-\eta_In^2}}{n^{2}}
\label{snscaleintermed}
\end{eqnarray}
Finally, for $L\gg \xi_z$, the integral over $x$ and $z$ in this expression converges as $L\rightarrow\infty$;
hence, in this regime, denoted region III in Fig. \ref{Fig5-TheoR}, all of the $S_n$ fall off with system size like ${1\over A}=L^{-2}$.
(Note that this regime must also have $L\ll L_{\rm c}$).
For $L\gg L_{\rm c}$ in the stable regime $D_{u x}>0$, denoted as region IV in Fig. \ref{Fig5-TheoR},
both contributions $\du1$ and $\du2$ to $\dut$ grow like $\ln\left({r\over a_l}\right)$ for large $r$, and hence, like
region I, this region will also exhibit algebraic decay of the $S_n$'s: $S_n\propto L^{-n^2\eta_{IV}} $,
albeit with $\eta_{IV}$ now taking on a different value:
\begin{eqnarray}\label{eta4}
\eta_{IV}={Cq_0^2\Upsilon\over8\pi}+
{q_0^2\Delta_u\over8\pi\sqrt{BD_{u x}}}\, .
\end{eqnarray}
However, since $\Delta_u$ is expected to be small, as discussed in Sec. \ref{App:UuApolar}, the numerical values of
$\eta_{IV}$ and $\eta_{I}$ are very close to each other, as shown in the simulation.

For $L\gg L_{\rm c}$ in the unstable regime $D_{u x}
<0$, denoted as region V in Fig.~\ref{Fig5-TheoR}, the system is unstable, and we expect dislocations
to proliferate, cutting off order parameter correlations at $L_{\rm c}$, and causing $S_n$ to again
decay like $L^{-2}$.

\section{\label{App:hydroP}Hydrodynamic theory predictions
for the polar active smectic P phase}

\subsection{\label{App:LinearPA}Linearized Eigenfrequencies and Instability threshold
%polar case
}

Fourier transforming Eqs. \eref{P_EOM_u},\eref{P_EOM_rho} in the main text and only keeping terms to linear order in $u$ and $\delta\rho$, we obtain
\begin{eqnarray}
\left[-i(\omega-gq_x^3)+\Gamma_u({\bf q})\right]u-iCq_z \delta\rho=f_u\,,
\label{uEOMFTpolar}\\
v_2q_xq_zu+\left[-i(\omega+v_\rho q_x)+\Gamma_\rho({\bf q})\right]\delta\rho=f_\rho\,,
\label{rhoEOMFTpolar}
\end{eqnarray}
with $\Gamma_u(\bf{q})$ and $\Gamma_\rho(\bf{q})$ having exactly the same expressions as in the apolar case.

Guessing that one of the eigenmodes has eigenfrequency
\begin{eqnarray}
\omega_1=-v_\rho q_x+\alpha({\bf q})
\end{eqnarray}
with $\alpha({\bf q}) =\cO(q^2)$, and inserting this guess into Eq. (\ref{uEOMFTpolar}) with $f_u$ set to zero on the right hand side,  we see that, to leading order in $q$, $u={Cq_z\over v_\rho q_x}\delta\rho$. Inserting this into equation (\ref{rhoEOMFTpolar}) with $f_\rho$ set to zero on the right hand side gives
\begin{eqnarray}
\alpha({\bf q}) =-i\Gamma_{\rho 2}({\bf q})\, ,
\end{eqnarray}
where we have defined
\begin{eqnarray}
\Gamma_{\rho 2}({\bf q})\equiv\Gamma_\rho({\bf q})+C{v_2\over v_\rho}q_z^2\, .
\end{eqnarray}
Thus,
\begin{eqnarray}
\omega_1=-v_\rho q_x-i\Gamma_{\rho 2}({\bf q})
\end{eqnarray}
Now looking for a second mode with eigenfrequency $\omega_2=\cO(q^2)$, Eq. (\ref{rhoEOMFTpolar}) with $f_\rho$ set to zero on the right hand side,  we see that, to leading order in $q$, $\delta\rho=- {iv_2\over v_\rho} q_z u$. Inserting this into equation (\ref{uEOMFTpolar}) gives
\begin{eqnarray}
\omega_2= gq_x^3-i\Gamma_{u 2}({\bf q})
\end{eqnarray}
where we have defined
\begin{eqnarray}
\Gamma_{u 2}({\bf q})\equiv\Gamma_u({\bf q})-C{v_2\over v_\rho}q_z^2=B{'''}q_z^2+D_{u x} q _x^2+Kq_x^4\, ,
\end{eqnarray}
with $B'''\equiv B-Cv_2/v_\rho$.  For $q_x\gg q_{\rm c}$, where the $D_{u x}$ term is negligible relative to the
$K$ term, $\Gamma_{u 2}({\bf q})$ again looks like an inverse smectic propagator, scaling like $1/q_z^2$ for $q_z\gg \lambda q_x^2$, and like $1/q_x^4$ for $q_z\ll \lambda q_x^2$.

Note that only $\omega_2$ involves the active tension $D_{u x}$. Hence, only this eigenfrequency can acquire a positive imaginary part, signaling an instability, when   the active tension $D_{u x}$ goes negative. And since the imaginary part of $\omega_2$ is manifestly an increasing function of $q_z^2$, it is obvious that this instability must first set in at $q_z=0$, $q_x=q_{\rm c}=\sqrt{-D_{u x}/K}$.

\subsection{\label{App:LinearPB}Linearized $u-u$ correlation functions
%polar case
}

With these eigenfrequencies in hand, we can now compute the spatiotemporally Fourier transformed $u-u$ correlation function in the linearized approximation precisely as we did in the apolar case. This gives
\begin{eqnarray}
	C_{uu}({\bf q}, \omega) 
&=&{\Delta_u((\omega+v_\rho q_x)^2+\Gamma_\rho^2({\bf q}))+(\Delta_{\rho x}q_ x^2+\Delta_{\rho z}q_z^2)
C^2 q_z^2\over(\omega-\omega_1)(\omega-\omega_2)(\omega-\omega_1^*)(\omega-\omega_2^*)}
\label{uqopolar}
\end{eqnarray}
This can again be integrated by parts to obtain the equal-time $u-u$ correlation function, albeit not as neatly as in the apolar case. We find, after considerably more algebra,
\begin{eqnarray}
C^{ET}_{uu}({\bf q})&\equiv&\langle |u({\bf q}, t)|^2 \rangle \nonumber \\
	&=&{\Delta_u\over 2\Gamma_{u 2}({\bf q})}+{\left[C^2 q_z^2\Delta_\rho({\bf q})+\Delta_u(\Gamma_{\rho}^2-\Gamma_{\rho 2}^2)\right]\left[\Gamma_{\rho 2}+\Gamma_{u 2}({\bf q})\right]\over2\Gamma_{u 2}({\bf q})
\Gamma_{\rho 2}({\bf q})\left[\left(\Gamma_{u 2}({\bf q})+\Gamma_{\rho 2}({\bf q})\right)^2+v_\rho^2 q_x^2\right]}\,.
\label{CETqpolar}
\end{eqnarray}

Note that once again, as in the apolar case, the first term will make a contribution to  $\langle|u({\bf r})|^2\rangle$ that is $\propto\sqrt{L}$ (for $L\ll L_{\rm c}$). Unlike the apolar case, however, the second term now only makes a finite contribution to $\langle|u({\bf r})|^2\rangle$. This can be seen by noting that for most directions of wavevector ${\bf q}$, the second term is independent of the magnitude $q$ of ${\bf q}$, since both the numerator and denominator of the second term scale like $q^6$ for such generic directions of ${\bf q}$.

This scaling does {\it not} hold, however, for $q_x\lesssim q_z^2$.
In this regime, the second term is readily seen to be $\propto {1/q^{2}}$. However, there is not enough phase space in the regime $q_x\lesssim q_z^2$  for this divergence to lead to any divergent fluctuations in real space; that is,
\begin{eqnarray}
\int_{1/L}^\Lambda dq_z \int_0^{q_z^2} dq_x {1\over q^2}\,,
\end{eqnarray}
does not diverge as $L\rightarrow\infty$.

The convergence of this second term means that its contribution to the
real space $u-u$ correlations can be dropped. We are thus left with only the first term in (\ref{CETqpolar}), which has exactly the same structure as the corresponding term in the apolar case. A moment's reflection reveals that this must mean that all of the scaling regimes found in the apolar case also exist in the polar case, in the linear approximation to the latter. The only difference is in region I, where the algebraic decay of order that was induced by the second term in the apolar case is absent in the polar one.

\section{\label{App:Nonlinear}Nonlinear effects for polar smectic P phases}

Going beyond the linearized approximation, the polar case has symmetry-allowed non-linear terms in its equation of motion
(i.e., those explicitly displayed in the equations of motion  \eref{P_EOM_u}, \eref{P_EOM_rho} in the main text). These become important only at the extremely
large length scale $L_{NL}$ discussed in the main text, and lead to the appearance of region VI in  Fig.~\ref{Fig5-TheoR}
 for the polar case. We have not been able to determine the behavior of the smectic order
 parameters in this regime, but strongly suspect, as discussed in the main text, that they fall off as $L^{-2}$ with increasing system size $L$ in this regime, which includes the $L\rightarrow\infty$ limit for the stable case.

The discussion in Secs. \ref{App:LinearPA} and \ref{App:LinearPB} was based entirely on the linear approximation to the full
equations
of motion (i.e., Eqs. (\ref{P_EOM_u}), (\ref{P_EOM_rho}) in the main text). As mentioned earlier, in the polar case the non-linearities in these equations are relevant
(technically, ``marginal") in $d=2$. These terms give rise to fluctuation-induced departures from the linearized theory  that
are proportional to the square of the noise strength $\sigma$  (note that our hydrodynamic  parameters $\Delta_{u, \rho x, \rho z}$ are all proportional to the {\it square} of the noise strength $\sigma$, as that noise strength is defined in equation (\ref{eq:angle_update}). To see this, note that, since the speed of the particles is fixed, and the angular noise in  (\ref{eq:angle_update}) is proportional to $\sigma$, then the {\it velocity} noise $f$ will be of order $\sigma v_0$, and, hence, also proportional to $\sigma$. Hence, the {\it mean-squared} forces,  which are proportional to the various $\Delta$'s, are proportional to $\sigma^2$.)
Furthermore,  in $d=2$,  they grow logarithmically with system size $L$. Hence,
for small noise strengths $\sigma$, the system must be larger than a nonlinear length scale $L_{N\!L}$ which grows very rapidly
with decreasing noise strength: $L_{N\!L}\sim\exp\left(\rm{constant}/\sigma^2\right)$ before the linear theory just described becomes
invalid.

What happens beyond this long length scale $L_{N\!L}$ remains an open question, which can only be answered by a full
renormalization group analysis. Due to the large number of marginal non-linearities (six) in our problem, such an analysis is
quite formidable,  and we have not attempted it. Here we will restrict ourselves to plausible speculation.

One possibility is that the non-linearities eventually induce the unbinding of dislocations, thereby making the smectic order parameter fall off far more rapidly than algebraically (presumably exponentially) for longer length scales. This would be consistent with our observation that, in the largest systems we study, dislocations do appear, and the smectic order parameter does fall off exponentially with increasing system size.

Independent of the question of dislocations, however, one thing that will certainly
happen for $L\gg L_{NL}$ is that the scaling law predicted by the linear theory relating higher harmonic smectic order parameters
$S_n$ to the lowest harmonic order parameter $S_1$, namely  $S_n \propto S_1^{n^2}$, will break down, since that scaling law
was derived in the linear approximation, in which the fluctuations of the layer displacement $\urt$ are Gaussian. Once non-linearities
become important (which, by definition, happens for $L\gg L_{NL}$), all of the fluctuations will become non-Gaussian, and this scaling
law will break down.

Therefore,  the hydrodynamic theory predicts that the  $S_n \propto S_1^{n^2}$ for $L\ll L_{NL}$, and that this law will break down for
$L\gg L_{NL}$. Since, in light of above discussion, we also expect correlations to fall off more rapidly for $L\gg L_{NL}$,  this means
we expect both a breakdown of scaling law $S_n \propto S_1^{n^2}$ relating the smectic order parameters, and the power law scaling
$S_1(L)\propto L^{-\eta_{I}}$ to fail at same system size $L$ (namely, $L\sim L_{NL}$). And this is exactly what we see in the
simulations, as illustrated in Figure \ref{Fig1-IsoRep}g of the main text, where we again plot $S_1(L)$, $S_2(L)^{1/4}$, and $S_3(L)^{1/9}$ versus system size $L$.
These lie on top of each other at small distances, consistent with the linear theory, but depart from each other at longer length scales,
signaling the onset of important non-linear effects at large distances. And at the system size $L$ where this happens, the order
parameters (particularly $S_1$), begin to depart from power law scaling with system size $L$ (see again Figure 1g of the main text). This is entirely
consistent with our argument above about the impact of the relevant nonlinearities on the order parameter correlations.  We therefore
interpret the system size at which both these phenomena happen as $L_{NL}$ for our system.

It would be a very dramatic illustration of this fundamental difference between the apolar and the polar case to show that
Eq. \eref{snscalen2} in the main text holds out to arbitrarily long length scales in the apolar case, which the theory says it should, since there
are no relevant nonlinearities allowed by symmetry in that case. However, this can only be tested in systems in which the
apolar case is actually stable (so one can explore region $IV$ in Figure 3). Alas, as discussed earlier, all of the apolar systems
we have investigated appear to be unstable.

%%%%%%%%%%%%%%%%%%%%%%%%%%%%%%%%%%%%%%%%%%%%%%%%%%%%%%%%%%%%%%%%%%%%%
\section{\label{App:oop}Orientational order parameters}
%%%%%%%%%%%%%%%%%%%%%%%%%%%%%%%%%%%%%%%%%%%%%%%%%%%%%%%%%%%%%%%%%%%%%

At low enough noise, the particles orientationally order in ways that reflect
the symmetry of their aligning interactions: for the the polar model,  the global polarity: $P = \left\vert \langle \textnormal{e}^{- \textnormal{i}\,\theta_j} \rangle_j \right\vert$ becomes nonzero.
For the apolar model, it is : $Q = \left\vert \langle  \textnormal{e}^{- \textnormal{i}\,2\theta_j} \rangle_j \right\vert $ that does so. In the polar model, the angle $\Phi$ denotes the orientation of the order, where the unit vector along the average direction of motion reads $\langle {\bf u}_i \rangle=(\cos\Phi, \sin\Phi)^T$. In the apolar model, an analogous angle can be defined for the nematic director; however, due to the  up-down symmetry in this case, its range is restricted to $\Phi\in[0,\pi]$.
\section{\label{App:smop}Numerical determination of smectic order parameters}
\begin{figure}
\begin{center}
\includegraphics[width=0.8\textwidth]{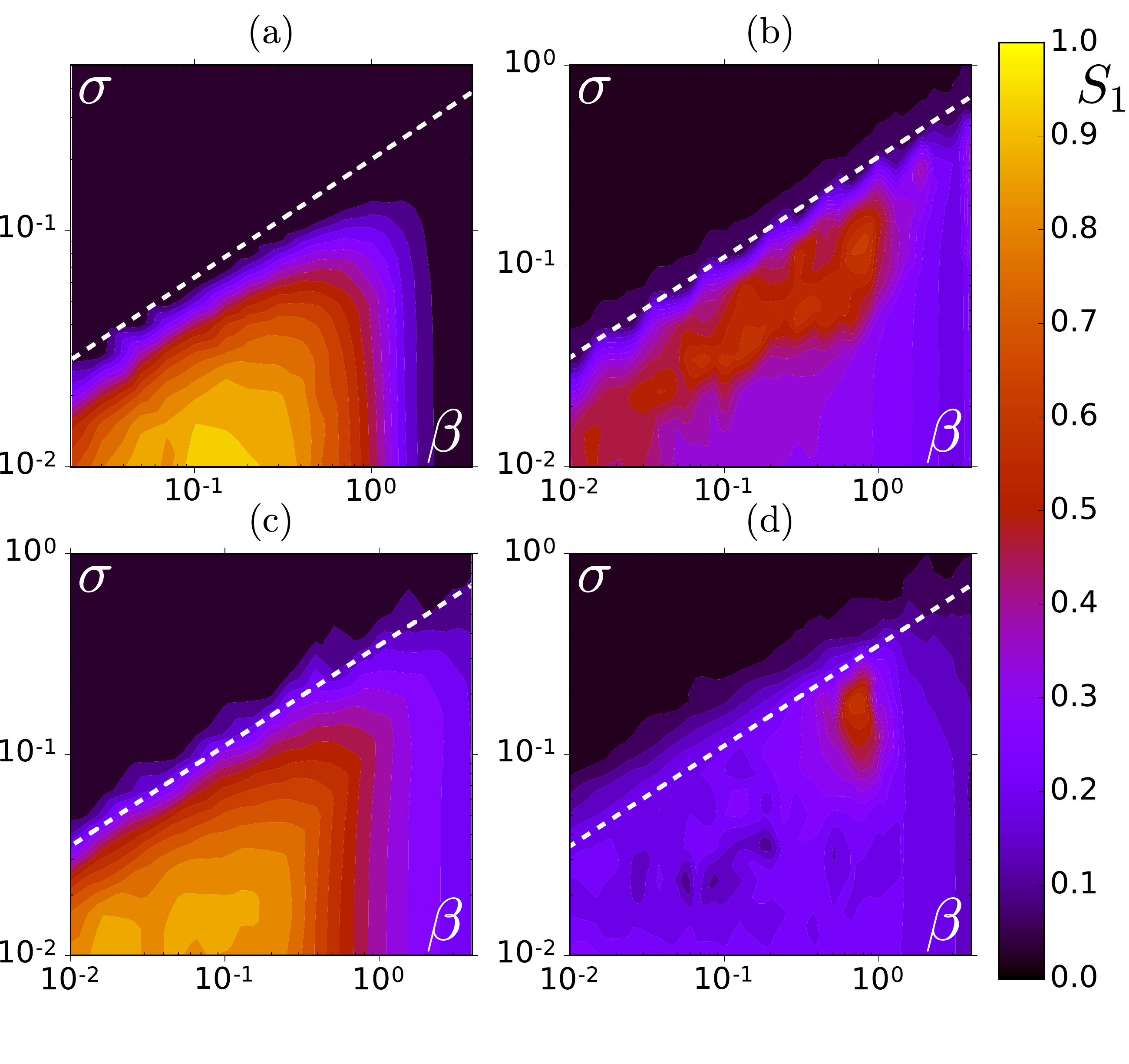}
\end{center}
\caption{
The smectic order parameter $S_1$ for small system sizes and different model variants plotted versus repulsion strength $\beta$ and angular noise $\sigma$. (a) Apolar model with nematic interactions and isotropic repulsion exhibiting smectic P configurations (Parameters: $L=20$, $\rho=8$, $v=0.3$). 
(b) Polar model with polar interactions and isotropic repulsion exhibiting smectic P configurations (Parameters: $L=24$, $\rho=10$, $v=0.2$). 
(c) Polar model with polar interactions and anisotropic repulsion with $\gamma=\pi/2$ (stronger repulsion to the sides) exhibiting smectic P configurations (Parameters: $L=24$, $\rho=10$, $v=0.2$). 
(d) Polar model with polar interactions and anisotropic repulsion with $\gamma=0$ (stronger repulsion to front and back) exhibiting smectic A (transient C) configurations (Parameters: $L=24$, $\rho=10$, $v=0.2$).
Initial condition: orientationally ordered system with no spatial order; Total simulation time: $T=2\cdot 10^6$. 
\label{figApp}
}
\end{figure}

The smectic phase corresponds to a periodic modulation of the density field in one direction of space; it is therefore useful to compute its Fourier transform. Numerically, we define a coarse grained number density field by binning the particles into a regular, rectangular grid of $M^2$ cells of size $\Delta r=L/M$, centered on the positions ${\bf r} \equiv (i_x\,\Delta r,i_y\,\Delta r)$, where the integers $(i_x,i_y) \in [0,M]\times[0,M]$. 
We now define $\rho({\bf r},t)$
as the number of particles in each cell divided by $\left(\Delta r\right)^2$. 
The normalized spatial Fourier transform can then be written in the compact form
\begin{eqnarray}
\hat \rho({\bf q},t) = \frac{\Delta r^2}{N} \sum_{i_x} \sum_{i_y} \rho ({\bf r},t)\,\textnormal{e}^{- i{\bf q}\cdot{\bf r} }
\label{ftdef}
\end{eqnarray}
 where $N$ is the total number of particles.  For ``perfect"  smectic order \cite{deGennes}, the   mean intensity $I({\bf q},t)=\langle |\hat \rho({\bf q},t)|^2\rangle$ would display in Fourier space a series of equally-spaced sharp Bragg  peaks  situated at wavevectors ${\bf q}_n = {{2\pi\,n}\over a_l}\, (\cos\Phi_{\rm S},\sin\Phi_{\rm S})^{\rm T}$, where $a_l$ is the mean layer spacing, the integer
$n\in[-\infty, \infty]$  is the peak index, and the angle $\Phi_{\rm S}$ gives the orientation of the mean  normal to the layers.
The trivial  central peak at ${\bf q}={\bf 0}$ is  normalized to one with our definition (\ref{ftdef}); the average height of the next peak defines the smectic order parameter: $S_1\equiv I({\bf q}_1,t)$. More generally, we can define an infinite series of smectic order parameters $S_n\equiv I({\bf q}_n,t)$.
For perfect smectic order, $S_n=1$ for all $n$, whereas the $S_n$  decrease with increasing $n$  in the presence of
fluctuations.

The value of $S_n$ depends on the coarse-graining scale $\Delta r$ used to define the density field.
We note that, decreasing $\Delta r$, $S_n(\Delta r)$ converges to some well-defined limit: $S_n(\Delta r) \approx S_n^0 -k_n\Delta r^{-2}$.
In practice, we used a sufficiently small, fixed $\Delta r=0.2$.  
We have also tested more elaborate techniques to estimate $S_n$ - e.g., using kernel density estimators or interpolations from different values of $\Delta r$ - to obtain $S_n$ in the limit $\Delta r \to 0$, to confirm that our main results on scaling relationships do not depend on these technical details.

In Fig \ref{figApp}{\cmag ,} we plot the first smectic order parameter $S_1$ for small systems as a function of the repulsion strength $\beta$ and noise intensity $\sigma$, which identifies regions with significant (local) smectic order.

%%%%%%%%%%%%%%%%%%%%%%%%%%%%%%%%%%%%%%%%%%%%%%%%%%%%%%%%%%%%%%%%%%%%%
\section{\label{App:params_repL}Numerical simulations: Matching layer spacing to system size}
%%%%%%%%%%%%%%%%%%%%%%%%%%%%%%%%%%%%%%%%%%%%%%%%%%%%%%%%%%%%%%%%%%%%%

It is well known from studies of equilibrium smectics that smectic order can be violently disturbed 
by even small departures of the system size from an integral multiple of the layer spacing 
\cite{deGennes}. Indeed, in some cases (specifically, when $\lambda\equiv\sqrt{K/B}<{a_l/4\pi}$, 
where $a_l$ is the layer spacing) such a mismatch can lead, in equilibrium, to an ``undulation 
instability" \cite{deGennes}. In order to minimize such effects, we have adjusted our numerical 
parameters whenever necessary to ensure that the observed smectic pattern in our simulations 
with periodic boundaries corresponds to the bulk behavior of active smectics. For numerical 
convenience,  we have accomplished this by adjusting the interaction range $r_{int}$ while 
keeping the system size fixed, until the average layer spacing a converged to unity. Only in the 
case of isotropic repulsion was this necessary, as only in that case was the average layer 
spacing significantly lower than the interaction range (specifically, we found $a_l
\approx0.95\,r_{int}$). The modified values for the interaction range needed to increase $a_l$ 
back to $1$ for isotropic repulsion were $r_{int} = 1.05$ for the F-model and $r_{int} = 1.06$ for 
the N-model. In order to keep the speed of individual particles comparable across different 
repulsion types, the speed (displacement per time step) was always measured in units of the 
interaction range.  In simulations, this choice of  $a_l=1$ 
 made it desirable, for the reasons just discussed, to  always work with systems of integer size $L$.  This not only minimized potential 
mismatch between the layer spacing and the system size, as discussed above, 
but also ensured that the spatial subdivision algorithm 
employed by us for efficient calculation of local interactions in large systems functioned 
correctly.

%%%%%%%%%%%%%%%%%%%%%%%%%%%%%%%%%%%%%%%%%%%%%%%%%%%%%%%%%%%%%%%%%%%%%
\section{\label{App:wind}Winding number}
%%%%%%%%%%%%%%%%%%%%%%%%%%%%%%%%%%%%%%%%%%%%%%%%%%%%%%%%%%%%%%%%%%%%%
Periodic boundary conditions make  
layered patterns with different integer ``winding numbers'' $w$ possible: $w=0$ means each layer
connects to itself at the periodic boundary, while for  $w=1$ each layer connects to the one above it, $w=2$ implies connecting to the layer two above it, etc. 
To minimize this artifact of periodic boundary conditions, we restrict our simulations of smectic patterns to ``flat" (i.e.,  $w=0$) configurations. However, it should be noted that in the apolar system,  the undulation instability at large length scales breaks the smectic layers, which may lead to the emergence of a smectic pattern with different winding numbers.

\ \vspace{1ex} \\

\end{document}